 \documentclass{aa}  

\usepackage{graphicx}
\usepackage{adjustbox}
\usepackage{amsmath}
\usepackage{txfonts}
\usepackage{hyperref}
\begin{document}

   \title{Refining pulsar radio emission due to streaming instabilities:
Linear theory and PIC simulations in a wide parameter range}

   \titlerunning{Refining pulsar radio emission due to streaming instabilities}

   \author{Alina~C.~Manthei 
          \inst{1} \fnmsep\thanks{Now at: Physikalisches Institut, Rheinische Friedrich-Wilhelms-Universität Bonn, 53115 Bonn, Germany},
          Jan Ben\'a\v{c}ek
          \inst{1},
          Patricio~A.~Mu\~noz
          \inst{1}
          \and
          J\"org~B\"uchner\inst{1,2}
          }
          
    \authorrunning{Manthei et al.}

   \institute{Center for Astronomy and Astrophysics, Technical University of Berlin, 10623 Berlin, Germany\\
              \email{manthei@physik.uni-bonn.de}
         \and
             Max Planck Institute for Solar System Research, 37077 G\"ottingen, Germany \\
             }

   \date{Received: ; accepted:}

 
  \abstract
   {Several important mechanisms that explain coherent pulsar radio emission rely on streaming (or beam) instabilities of the relativistic pair plasma in a pulsar magnetosphere.
    However, it is still not clear whether the streaming instability by itself is sufficient to explain the observed coherent radio emission.
    Due to the relativistic conditions that are present in the pulsar magnetosphere, kinetic instabilities could be quenched.
    Moreover, uncertainties regarding specific model-dependent parameters impede conclusions concerning this question. }
   {We aim to constrain the possible parameter range for which a streaming instability could lead to pulsar radio emission, focusing on the transition between strong and weak beam models, beam drift speed, and temperature dependence of the beam and background plasma components.}
   {We solve a linear relativistic kinetic dispersion relation appropriate for pulsar conditions in a more general way than in previous studies, considering a wider parameter range.
   	In doing so, we provide a theoretical prediction of maximum and integrated 
   	growth rates as well as of the fractional bandwidth of 
   	the most unstable waves for the investigated parameter ranges.
   The analytical results are validated by comparison with relativistic kinetic particle-in-cell (PIC) numerical simulations}
   {We obtain growth rates as a function of background and 
   	beam densities, temperatures, and streaming velocities while finding a remarkable agreement of the linear dispersion predictions and
   	numerical simulation results in a wide parameter range.
   	Monotonous growth is found when increasing the beam-to-background density ratio.
   	With growing beam velocity, the growth rates firstly increase,
   	reach a maximum and decrease again for higher 
   	beam velocities.
   	A monotonous dependence on the plasma temperatures is
   	found, manifesting in an asymptotic behaviour when reaching colder temperatures.
   	A simultaneous change of both temperatures proves not to be a mere linear superposition of both individual temperature dependences.
   	We show that the 
   	generated waves are phase-coherent by calculating the fractional bandwidth. 
    }
   {Plasma streaming instabilities of the pulsar pair plasma can
   	efficiently generate coherent radio signals
   	if the streaming velocity is ultra-relativistic with
   	Lorentz factors in the range $13 < \gamma < 300$,
   	if the background and beam temperatures are small enough 
   	(inverse temperatures $\rho_0 ; \rho_1 \geq 1$, 
   	i.e. $T_0 ;  T_1 \leq  6 \times 10^9$), 
   	and if the beam-to-background plasma density ratio 
   	$n_{1} / (\gamma_\mathrm{b} n_{0})$ exceeds 
   	$10^{-3}$, which means that $n_{1} / n_{0}$ has to be between 1.3 and 20\% (depending on the streaming velocity) 
   }


   \maketitle
%

\section{Introduction}
\label{sec:intro}
Since the discovery of radio pulsars more than half a century ago in 1967 ~\citep[see][]{Hewish1968}, 
a large number of physical models have been proposed to explain their extremely coherent pulsar radio signals
 \citep{Goldreich1969,Ruderman1975,Blaskiewicz1991,Kramer2002,Petrova2009,Beskin2018,Liu2019,
Philippov2020}.
Nevertheless, the scientific community has not yet come to an agreement on a mechanism
that can fully explain the observations \citep{Melrose2017c,Beskin2018}. 

The current standard model for the (pair) plasma in the pulsar magnetosphere can be summarised as follows \citep{Goldreich1969,Sturrock1971,Ruderman1975}: Due to the fast rotation (with periods from milliseconds to seconds) and the strong magnetic field of the spinning neutron star
(typically $10^{11}-10^{13}\;\;{\rm gauss}$), an electric field has to be generated in order 
to satisfy the ideal frozen-in condition on its surface.
This electric field features a very large electrical
conductivity and is directed mostly parallel to the magnetic field lines.
The screening of this electric field by the plasma already existing in the magnetosphere causes an induced polarisation electric field, forcing the aforementioned plasma to co-rotate with the neutron star.
The charge density associated with this induced electric field (and associated co-rotation potential) is called the
co-rotation charge density or the Goldreich-Julian particle density. It typically amounts to $10^{11}-10^{12}\;\;{\rm particles/cm^3}$ and decreases rapidly with height as far as
the so-called light cylinder, where the co-rotation speed of the plasma reaches the speed of light.
The light cylinder boundary determines the characteristic size of a pulsar magnetosphere.
In the polar cap region of the pulsar magnetosphere, the magnetic field lines are open 
such that the plasma does not co-rotate there, but rather particles  
are free to stream and escape along the magnetic field lines.
Near the neutron star surface at the polar caps, this escape of particles causes a local plasma depletion on a length scale in the order of $10^4{\rm cm}$.
It leads to a vacuum with an associated strong parallel electric field $\vec{E}\cdot\vec{B}\neq0$ and an associated potential of the order of $10^{12}$~V.
This vacuum region is often called a `gap', resembling an electrostatic double layer.
We note that outside of this gap, $\vec{E}\cdot\vec{B}=0$ almost everywhere in 
the pulsar magnetosphere, such that particles in that region are not accelerated by this mechanism.
The charge of the escaping particles depends on the sign of the co-rotation charge;
if it is positive near the polar caps, positrons will be accelerated
outwards from the neutron star surface while electrons will move in the opposite direction, towards the neutron star.
The escaping particles moving along the curved magnetic field lines form the so-called `primary beam'.
They have ultra-relativistic energies and relativistic Lorentz factors $\gamma = (1-\beta^2)^{-1/2}$ (where 
$\beta = v/c$) of the order of $\gamma\sim10^6-10^7$.
This eventually triggers the emission of curvature radiation photons at
$\gamma$-ray frequencies.
Those photons subsequently undergo pair production, producing 
a plasma out of electrons and positrons \citep{Sturrock1971,Cheng1977b}.
Assuming the sign of the co-rotation charge as above, the newly created positrons move away from the neutron star, while
the newly created electrons move towards it.
These more recently produced particles are called the secondary plasma population, or `secondary beam'. According to the estimations of \citet{Ursov1988}, these particles feature relativistic Lorentz factors in the range $\gamma_\mathrm{min}\sim10$ to $\gamma_\mathrm{max}\sim 10^3-10^4$.

Each of these particles is also prone to further emission of $\gamma$-photons
due to curvature radiation,
followed by a subsequent pair-creation event, at different heights in the gap region.
Moreover, synchrotron radiation (due to particle gyromotion in the strong magnetic field) emitted by the secondary beam also contributes to the increasing number of electron-positron pairs since these energetic photons can also undergo pair production.
This chain of events leads to the periodic breakdown of the gap vacuum by the generated
electrons and positrons when their charge density reaches the Goldreich-Julian density \citep{Goldreich1969}.
Due to the reduction of the acceleration electric potential by plasma screening in the neighbourhood of the pair production event, particles are no longer accelerated and consequently vacate the region.
Since this also implies that no further pair production occurs, the gap region is empty
thereafter, and the process of filling it with particles until the renewed attainment of the Goldreich-Julian density may re-start.  This periodic `sparking' process occurs on timescales of microseconds \citep{Ruderman1975}.

The aforementioned curvature radiation itself, however, cannot be accountable for the observed coherent radio pulses. 
Coherent radiation means that all particles radiate in phase with each other.
An investigation of observational data from six pulsars testing this hypothesis was carried out by \citet{Lesch98} who discarded it.
The observational finding can be reinforced by theoretical arguments.
With respect to the emission by the primary beam, the wavelengths of its $\gamma$-ray photons are much smaller than the mean spacing between the plasma particles, such that the curvature radiation by the primary beam is incoherent radiation at much higher frequencies with respect to the radio range.
Curvature radiation by the secondary beam, however, was one of the first proposed models 
to explain pulsar radio emission~\citep{Ruderman1975}.
Owing to its lower energy and higher density relative to the primary beam (by a factor of $10^3-10^5$),
this beam emits curvature photons with wavelengths much larger than the mean particle spacing,
possibly favouring coherent radiation~\citep{Beskin1993}.
In this case, the plasma frequency and emitted curvature radiation would indeed fall in the radio range.
Yet, later theoretical arguments have shown that this hypothesis suffers
some deficiencies as well. 
One of the most important is that it cannot
lead to emission with brightness temperatures beyond $> 10^{13}$~K~\citep[Appendix B]{Melrose1978}.
Hence, this emission is insufficient to explain the observed brightness temperatures
of pulsar radio emission, which may reach up to $10^{41}$~K~\citep{Hankins2007}.
The observed emission can only be explained by a coherent mechanism that is characterised by a high brightness temperature~\citep{Eilek2016}.

Another possible source for the pulsar radio emission consists in a beam instability (also known as streaming or two-stream instability).
In this case, the source of free energy is rooted in the relative flow between populations of particles that move at different (mean) velocities.

Kinetic instabilities may generate unstable plasma waves that can be transformed into electromagnetic waves and eventually be observed in the radio regime
\citep{Melrose1999}. At least three scenarios are conceivable for the generation of this (two-)stream instability \citep{Asseo1998}:

Firstly, an interaction between the primary beam and the secondary plasma is possible \citep{Ruderman1975,Cheng1977c,Buschauer1977,Arons1981}; however, this process has already been found to be inefficient and cannot account for the generation of the observed radiation. 
One of the main arguments against it consists in the lack of time for the instability to develop. 
\citet{Usov1987} pointed out that a time period of $\tau_\mathrm{I}\sim10^{-4}(r/R)^{\frac{3}{2}}$~s is necessary for the resulting instability development, assuming a distance $r$ from a pulsar with radius $R$ and provided the parameters of the model in \citet{Ruderman1975}. The plasma particles, by contrast, travel the distance $r$ within $\tau_0\sim3\cdot 10^{-5} (r/R)$~s, which is smaller than $\tau_\mathrm{I}$ for any $r>R$.

Secondly, a different model hypotheses counter-streaming electron and positron populations to be responsible for the observed coherent radio emission \citep[e.g.][]{Ruderman1975,Cheng1977b,Weatherall1994}.
A relative drift between both populations may be caused by the fact that the secondary plasma is exposed to an electric field that arises from the magnetic field curvature and 
the Goldreich-Julian charge density condition \citep{Cheng1977c}.
However, this kind of instability could easily be quenched by the 
relativistic temperature of the pulsar magnetosphere plasma~\citep{Buschauer1977}.
Thus, this type of two-stream instability is not likely to be the origin of pulsar radio emission either.

The third possibility, the one that is further investigated in this work, consists of an interaction between two successively emitted plasma bunches, composed of both electrons and positrons. 
These bunches occur due to the non-stationarity of the plasma emission mechanism in the polar-cap regions, the aforementioned `spark' events \citep{Ruderman1975,Cheng1977b}.
As the filling with new electron-positron pairs takes some time, the previous bunch has already travelled a certain distance outwards when the next one is emitted. 
The fast particles from the freshly emitted bunch may eventually catch up with the slow ones from the prior particle bunch \citep{Asseo1998}.
The interaction between the particles from the new bunch, forming a particle beam, and those from the former one, considered as the background plasma, can lead to the development of a beam instability, a process proposed in \citet{Usov1987} \citep[see also][]{Ursov1988,Usov2002}. 
This process underlies the recent studies by~\citet{Rafat2019b} whose work we extend and generalise to
a broader parameter range.

The pulsar pair plasma, generated by the pair creation process, is relativistic. 
A relativistically covariant way to take these effects into account is using one-dimensional Maxwell-J\"uttner \citep{Juttner1911} distributions for both the beam and the background plasma \citep{Asseo1998}. 
The assumption of one-dimensionality 
can be adopted in the lower pulsar magnetosphere due to the dominant magnetic field, implying that the drift perpendicular to the magnetic field does not significantly contribute to the particle motion\footnote{We note that, by contrast, when the emitted radiation approaches the light cylinder, cyclotron effects might contribute since the electron cyclotron frequency and electron plasma frequency are of the same order of magnitude in this region. 
Thus, only one-dimensional analysis is no longer adequate in such a case \citep{Luo2001}.} By means of numerical simulations of the pair creation process,  
\citet{Arendt2002} found that the typical relativistic
bulk flow speed is in the range $p/mc \approx 32-178$  for a magnetic field of $B_* = 10^{12}$~gauss.
The inverse momentum width $p/\Delta p \approx 0.23-0.79$  leads to relativistic  temperatures of the particle bunches.
They also found that the resulting distribution function is very close to the Maxwell-Jüttner distribution.

It is still unclear, however, what (relativistic) temperatures should be considered. 
The various pulsar models provide different assumptions here.
\citet{Weatherall1994} assume the temperature of the beam to be non-relativistic, while the one of the background is supposed to be highly relativistic in this model. 
In the case of \citet{Rafat2019b}, the same temperature is used for both components.
At the distance from the pulsar where the interaction of two bunches takes place, as proposed in the model by \citet{Usov1987}, the temperatures of both plasma populations remain unspecified.

Within the same model by \citet{Usov1987}, however, it might be possible that the electrons and positrons from a previously emitted bunch of particles could have adiabatically cooled down by the time the fast particles from the succeeding bunch reach them.
Choosing the rest frame of the secondly emitted, still hot bunch, that is therefore to be regarded as the background plasma, the firstly emitted and already cooled-down one moves with a relative velocity with respect to the background and can be considered as the beam. 
A lower beam temperature could lead to a stronger instability compared to the case in which both populations are hot.

It is also conceivable that the temperature has already dropped from the neutron star surface to the interaction region, such that the background and the beam equally face a cool-down, while the ratio of both temperatures persists.

Since the model by \citet{Usov1987} is based on a two-stream instability due to interaction between a previously emitted bunch and the successive one, the densities of the beam and background plasma are assumed to be approximately equal. 
\citet{Rafat2019b}, following \citet{Egorenkov1984}, by contrast, conjecture a density ratio of $n_{1}/n_{0}=10^{-3}$ in the background reference frame to be appropriate  (corresponding to a weak beam), which makes an investigation of the influence of different densities worthwhile\footnote{We note that in the following, we use the symbols $n_0$ and $n_1$ to describe the densities of beam and background in their respective reference frames.}.

Another important parameter whose influence on the streaming instability has to be understood
more thoroughly is the relative velocity of the of beam and background plasma. The exact relative velocity of particles between two bunches, determining the relative drift between beam and background,
is not fixed to an explicit value by the presumed model of \citet{Usov1987}.
We note that the beam-background interaction could take place not only between the fastest particles from the new bunch and the slowest ones from the previously emitted one, but also between several groups of particles with intermediate velocities \citep{Asseo1998}. Thus, the range of possible relative drift speeds between the beam and background plasma can be very broad. 
Moreover, the more the beam drift speed is reduced, the closer the system gets to a strong-beam model that is defined by $r_\mathrm{n} = n_{1} / (\gamma_\mathrm{b}n_{0}) \rightarrow 1$.
In our analysis, we orient ourselves by numerical experiments of \citet{Arendt2002}
as well as by the parameters
used in the analytical relativistic dispersion relations investigated by \cite{Rafat2019b}.

Instabilities of relativistic pair plasma beams have been already studied using 
kinetic plasma simulations
\citep{Silva2003,Tautz2007,Cottrill2008,Lopez2014,DAngelo2015,Lopez2015,
Shukla2015}.
Only a few studies, however,
have assumed relativistic covariant Maxwell-J\"uttner distributions, and if so, they were
mostly considered to be three-dimensional  \citep{Bret2008,Bret2010}.
Only \citet{Shalaby2017} regarded a one-dimensional Maxwell-J\"uttner  distribution as more appropriate under the strongly magnetised pulsar conditions. They used it
in order to investigate a
relativistic two-beam instability with relativistic beam Lorentz factors of $\gamma_\mathrm{b} \leq 4$.
Beyond that, following up on these results, \citet{Shalaby2018} also analysed higher beam velocities ($\gamma_\mathrm{b} = 100$) for an inhomogeneous background.

A discussion of the influence of the adopting this Maxwell-J\"uttner  distribution function and its associated relativistic effects on the wave dispersion, with a view to pulsar radio emission, has already been carried out by \citet{Rafat2019b}. 
Their investigations led to the conclusion that the above mentioned beam-driven instabilities do not suffice as a primary source for the observed pulsar radiation.
However, this might be due to the limited range range of plasma parameters for which \citet{Rafat2019b} carried out their calculations.

Beyond that, \citet{Rahaman2020} recently solved a hot relativistic dispersion relation appropriate for a pulsar plasma magnetosphere as well. Similarly to our study, they also considered the streaming instability due to the overlap of successively emitted plasma bunches within the \citet{Usov1987} model, but additionally taking the pulsar magnetosphere geometry into account and comparing among several pulsar emission models. 
They found the most efficient mechanism that should be responsible for the highest growth rates
among those models and the expected emission height.
Different from our work, however, they did not carry out numerical simulations to confirm the solutions of their dispersion relations.

Here, we re-investigate the question whether streaming instabilities may cause the observed
pulsar radio emission by extending the investigations by \citet{Rafat2019b} and their range of considered plasma parameters.
As far as we know, for the first time, we investigate  the dependence of 
wave dispersion branches and instability growth rates on a wide range of plasma properties,
in order to obtain stronger constraints on the possible conditions
for an efficient radio-pulsar emission due to streaming instabilities.
Those investigated parameters include the temperature of the background and beam, beam velocity, and the beam-background density ratio.
Beyond the scope of previous studies, we additionally compare our analytical linear dispersion relation solutions
with suitable kinetic particle-in-cell (PIC) simulations.

The outline of this paper is the following. 
First, we give a short summary of the theoretical basis (Sect.~\ref{sec:theory}). Then, we state the methods used for solving the dispersion relations, and we describe our simulation setup (Sect.~\ref{sec:methods}).
Next, we conduct several parametric studies in order to constrain the parameter regime where instabilities could be able to grow and generate plasma waves (Sect.~\ref{sec:results}).
In Sect.~\ref{sec:discussion}, we compare the analytical dispersion relation solutions and the kinetic simulation results.

\section{Dispersion relation for a Maxwell-J\"uttner velocity distribution in a pulsar pair plasma}
\label{sec:theory}

In order to cover a wide range of plasma parameters, this section is devoted to the derivation of an appropriate
generalised kinetic dispersion relation based on the model by \citet{Ursov1988}. Here we consider a pair plasma composed of electrons with density $n_{-}$
and positrons with density $n_{+}$. Within the bunch model by \citet{Ursov1988},
we assume a `background'  (denoted with subscript 0) at rest and a `beam' (denoted with subscript 1)  drifting outwards parallel to the magnetic field at a relative speed of $v_\mathrm{b}=\beta_\mathrm{b} c$.
We note that we do not intend to investigate instabilities caused by the relative streaming of
both populations due to the magnetic field curvature~\citep{Cheng1977c}.
Hence we neglect the small difference between $n_{-}$ and $n_{+}$ that the latter would cause.
Instead, within the model by \citet{Ursov1988},
each background and beam plasma population contains the 
same number of electrons and positrons. 
Therefore, they would simply be equal to $n_0$ in the reference frame of the background plasma.
Similarly, the electrons and positrons in the beam
feature a density $n_1$ in their own reference frame. 
All our calculations are performed in the reference frame of the background
such that the beam density is equal to $n_{1}/\gamma_\mathrm{b}$ in this reference frame, with the 
relativistic Lorentz factor $\gamma_\mathrm{b}=(1-\beta_\mathrm{b}^2)^{-1/2}$. 
Such a plasma system can be described by different distribution functions.
At the beginning, the discussion below does not assume a specific
form of the distribution function.
The specific relativistic plasma dispersion function that is used in our study is described after that.

Generally, the linear plasma response to small perturbations (linear dispersion theory) can be described by the
dielectric tensor $K_{ij}(\omega, \vec{k})$.
By equating its determinant to zero,
a dispersion relation $\omega=\omega(\vec{k})$ can be obtained that determines 
all the properties of linear plasma electrostatic wave modes.
In general, both frequency $\omega=\omega_\mathrm{r}+i\omega_\mathrm{i}$ and wave vector $\vec{k}=\vec{k}_\mathrm{r}+i \vec{k}_\mathrm{i}$ can be complex numbers.
The question whether the wave growth is chosen to be considered as purely temporal or purely spatial is determined by the specific circumstances and also by practical aspects of the particular application. 
Depending on the physical problem under consideration, the dielectric tensor might be simplified by appropriate
choices \citep{Sturrock1958,Weatherall1994}, facilitating the calculations.
For our purpose in this work, the frequency $\omega$ is chosen to be imaginary while keeping the wavenumber $k$ real. 
This implies that $\omega_\mathrm{i}$ characterises the wave growth.
For $\omega_\mathrm{i}>0$, the  wave grows exponentially, being the hallmark of a plasma instability, whereas the wave is damped for $\omega_\mathrm{i}<0$.

Following \citet{Gedalin1998,Melrose1999}, we calculate 
the dielectric tensor $K_{ij}$ in a three-dimensional Cartesian geometry, assuming a background magnetic field oriented in the
direction denoted as $3$.
We further assume that both wave propagation and field fluctuations only occur in direction parallel to the magnetic field, corresponding to a one-dimensional plasma model.
This has proved an appropriate approximation
to describe streaming instabilities in a pair plasma in the pulsar magnetosphere emission region,
mainly justified by the strong magnetisation in this region \citep{Weatherall1994}.
Furthermore, the wave frequency is assumed to be much smaller than the corresponding cyclotron frequency, meaning $\omega\ll\Omega_\mathrm{e^{\pm}}/\langle \gamma \rangle$ with $\Omega_\mathrm{e^\pm}=eB/m_\mathrm{e}$.
Hence cyclotron resonance effects can be neglected.

Under the aforementioned assumptions, the dispersion relation simplifies considerably.
Only the $K_{33}$ component parallel to the magnetic field (out of six independent components) does not vanish.
Thereby, the full dispersion relation $|K_{ij}(\omega,k)|=0$ becomes
\begin{equation}
\label{eq:dispersion}
K_{33}=1-\frac{\omega_\mathrm{p}^2}{\omega^2}\cdot z^2 W(z)=0,
\end{equation}
where $\omega_\mathrm{p}$ is the plasma frequency of both populations, including a relativistic correction due to the beam Lorentz factor:
\begin{equation}
\omega_\mathrm{p}^2 = \frac{e^2}{\epsilon_0 m_\mathrm{e}}\left( 2n_0 + 2\frac{n_1}{\gamma_b} \right).
\end{equation}
In the above equation, $e$ denotes the elementary charge, $\epsilon_0$ stands for the vacuum electric permittivity, and $m_\mathrm{e}$ is the electron (or positron) mass.
We note that $\omega_\mathrm{p}$ is the physical fundamental oscillation frequency in a pair plasma,
different from $\omega_\mathrm{pe}=e^2n_{-}/(m_\mathrm{e}\epsilon_{0})= \omega_\mathrm{p}/\sqrt{2}$ with $n_- = n_0 + n_1$ \citep{Stenson2017}. 
The parameter $z=\omega/(k c)$ is used to indicate the phase velocity of a wave with frequency $\omega$ and wavenumber $k$.
The so-called relativistic plasma dispersion function $W(z)$ in Eq.~\eqref{eq:dispersion} is defined as
\begin{equation*}
W(z)=W_{0}(z)+r_\mathrm{n}\gamma_\mathrm{b}W_{1}(z),
\end{equation*}
where $r_\mathrm{n}=n_{1}/\left(\gamma_\mathrm{b} n_{0}\right)$ describes the ratio of densities between beam and background plasma in the rest frame of the background.
The respective background ($W_{0}$) and beam ($W_1$) relativistic dispersion functions can be written in the following way:
\begin{multline}\label{eq:rel_function}
W_{\alpha}(z) = \frac{1}{n_\alpha}\int_{-\infty}^{\infty}\mathrm{d}u\frac{\mathrm{d}g_{\alpha}(u)/\mathrm{d}u}{\beta-z}=\frac{1}{n_\alpha}\int_{-1}^{1}\mathrm{d}\beta\frac{\mathrm{d}g_{\alpha}(u)/\mathrm{d}\beta}{\beta-z}, \\ \\ ~\alpha = 0,1, \qquad
\end{multline}
where $u = \beta / (1 - \beta^2)^{\frac{1}{2}}$.
This transcendental function 
has a singularity at $\beta=z$ that
has to be taken care of by applying Cauchy's integral formula. 
Further properties of this function are discussed in \citet{Godfrey1975a,Rafat2019b}.

The relativistic function (Eq.~\eqref{eq:rel_function}) and so the dispersion relation (Eq.~\eqref{eq:dispersion}) require the specification of an
appropriate distribution function $g_{\alpha}$.
In order to account for possible relativistic temperatures as well as relativistic bulk flow velocities of both beam and background plasma, we model their velocity distribution functions $g_{\alpha}$ as
relativistically covariant Maxwell-J\"uttner distributions. 
These velocity distributions depend on both the inverse temperature in units of the particle rest energy, denoted as $\rho=m_\mathrm{e} c^2/(k_\mathrm{B}T)$, as well as on the relativistic Lorentz factor $\gamma=(1-\beta^2)^{-1/2}$.
This way, plasmas with relativistic temperatures are characterised by $\rho<1$ while relativistic plasma flows have the property $\gamma\gg1$.

The resulting Maxwell-J\"uttner velocity distribution function
of the entire plasma is composed of the individual distributions for each plasma population. 
In the reference frame of the background, it can be written as \citep{Rafat2019b}:
\begin{align}
g(u)&=g_{0}(u)+g_{1}(u)\label{reldisp-total}\\
g_{0}(u)&=n_{0}\cdot\frac{\exp\left(-\rho_{0}\gamma\right)}{2K_{1}(\rho_0)},\label{reldisp-mj0}\\
g_{1}(u)&=\frac{n_{1}}{\gamma_\mathrm{b}}\cdot\frac{\exp\left(-\rho_{1}\gamma_\mathrm{b}\gamma\left(1-\beta\beta_\mathrm{b}\right)\right)}{2K_{1}(\rho_1)} \label{reldisp-mj1}\text{,}
\end{align}
where 
$K_{1}$ denotes the Macdonald function of first order (also known as modified Bessel function of the second kind).

The source of free energy for kinetic streaming instabilities can be traced back to
a positive slope $\mathrm{d}g(u)/\mathrm{d}u>0$ for $u>0$ of the 
joint velocity distribution of Eq.~\eqref{reldisp-total} \citep{Baum1997}.
Since there are more particles with higher velocities than the wave phase speed in the positive 
slope region, a wave with $\beta<\beta_\mathrm{b}$ is able to gain more energy than it loses by interactions with lower-momentum particles. 
Such a positive slope is possible if the centre (bulk flow speed) of the beam distribution \eqref{reldisp-mj1} in velocity space is sufficiently separated from the background distribution function \eqref{reldisp-mj0}.
This implies the existence of a well-defined minimum in the joint velocity distribution \eqref{reldisp-total}. This condition can be formally stated
by means of the so-called Penrose criterion, also known as `separation condition' (between the beam and background distributions). In the model presented here, it predicts the following minimum beam drift speed that would be necessary for the system to become unstable \citep{Melrose1986,Rafat2019b}:
\begin{equation}
\gamma_\mathrm{b,min} \approx 7.8 \left(\frac{1}{r_\mathrm{n} r_\rho r_\mathrm{K}} \right)^{0.076} \left(\frac{1}{\rho_0}\right)^{1.07},  
\label{eq:penrose}
\end{equation}
\begin{equation}
\qquad r_\rho = \frac{\rho_1}{\rho_0}, \qquad r_\mathrm{K} = \frac{K_1(\rho_0)}{K_1(\rho_1)}.
\end{equation}

Besides the growth rates, corresponding to the imaginary parts of the frequency, the fractional bandwidth of the growing modes can be examined by studying the real part of the frequency.
The bandwidth is defined as the frequency range between the two values of $\omega_\mathrm{r}$ for which the growth rate ($\omega_\mathrm{i}$) has decreased to half of its maximum value (FWHM).
To obtain a dimensionless quantity, a division by $\omega_\mathrm{r}$, where the maximum of $\omega_\mathrm{i}$ occurs, is performed. 
This yields the fractional bandwidth $\Delta \omega_\mathrm{r} / \omega_\mathrm{r}\left(\omega_\mathrm{i,max}\right)$.
This quantity determines the maximum value of the growth rate for which the random phase mixing approximation is valid. If the growth rate value exceeds the value of the fractional bandwidth, the unstable wave mode is phase-coherent.

One way to quantify the efficiency of the wave growth is via the growth factor $G$,
a quantity that, in the case of purely temporal growth, is defined as \citep{Melrose1986}:
\begin{equation}\label{eq:def_g}
G=\omega_\mathrm{i} \Delta t,
\end{equation}
where $\Delta t$ is the growth time. Within the \citet{Ruderman1975} sparking model, it can be interpreted as the time span between two pair creation events. 
The boundary between efficient and inefficient growth occurs at $G=1$.
Thus, a necessary condition for the instability to be effective is $G>1$.

\section{Methods}
\label{sec:methods}
The linear kinetic dispersion relation of Eq.~\eqref{eq:dispersion} can be solved numerically,
and its solutions can be compared to results of particle-in-cell (PIC) code simulations. 
Both methods are described in this section.

\subsection{Numerical solution the dispersion relation}
\label{subsec:num}

Our aim is to solve the linear kinetic dispersion relation \eqref{eq:dispersion} in a wider range of
plasma parameters and
by assuming fewer simplifications than previous studies (cf.~\citet{Rafat2019b}).
\citet{Rafat2019b} found the roots of the dispersion relation ~\eqref{eq:dispersion} by
using an expansion of $K_{33}$ about $\omega_\mathrm{r}$ and $k_\mathrm{r}$ to the first order in $\omega_\mathrm{i}$ and $k_\mathrm{i}$.
This approach assumes $\omega_\mathrm{i}/\omega_\mathrm{r}\ll1$ and $k_\mathrm{i}/k_\mathrm{r}\ll1$.
In contrast, we do not make such an assumption, allowing for arbitrary
values of $\omega_\mathrm{i}$.
To do so, first we sample independently both real and imaginary part of the complex function $K_{33}$ on a grid in the plane $(\omega_\mathrm{r},\omega_\mathrm{i})$.
This sampling is carried out for a given $k_0$ where we expect to find 
a root of the dispersion relation.
Then, we trace the
contour levels of the real and imaginary parts of $K_{33}$ that are  equal to 0 for the chosen $k_0$.
The solutions of Eq.~\eqref{eq:dispersion} will thus be given 
by the intersections of those contours where both real and imaginary parts vanish simultaneously.
A first-order approximation for those intersections is found by a 
geometrical method based on determining the intersections of the piecewise straight lines joining each grid point of the `0'-contour levels discretised onto the grid $(\omega_\mathrm{r},\omega_\mathrm{i})$.
This method leads to intersections that are accurate above the grid resolution level $\Delta \omega_\mathrm{r}=(\omega_\mathrm{r,max}-\omega_\mathrm{r,min})/(n_\omega-1)$, $\Delta \omega_\mathrm{i}=(\omega_\mathrm{i, max}-\omega_\mathrm{i,min})/(n_\omega-1)$, where $n_\omega=50$, for an initially given $\omega_\mathrm{r/i, min}$ and $\omega_\mathrm{r/i,max}$.
Subsequently, a high-accuracy intersection is found by using the previous initial intersection 
as an initial guess for the multi-dimensional root finding function named \texttt{root} of the module \texttt{optimise}
in the Python library \texttt{SciPy}, which uses the non-linear least squares Levenberg Marquardt algorithm \citep{Levenberg1944, Marquardt1963}.  This algorithm combines the Gauss-Newton algorithm and the method of gradient descent.
Its tolerance for convergence amounts to $10^{-12}$.
If this root finding algorithm fails to converge, a different approach is used:
the modified Powell's hybrid method \citep{Powell1964} with a tolerance for convergence of $10^{-6}$.
The intersection that is eventually discovered, $(\omega_\mathrm{r0},\omega_\mathrm{i0}, k_0)$, represents a root of Eq.~\eqref{eq:dispersion}.
Once this initial intersection has been found, a solution for the 
next point $k_1=k_0 + \Delta k$ with $\Delta k = (k_\mathrm{max}-k_\mathrm{min})/(n_k-1)$, $n_k=2000$ (given the initially set range of $k_\mathrm{max}-k_\mathrm{min}$) is looked for by using as initial guess 
a linear extrapolation (adapted to the resolution in $k$) of  $(\omega_\mathrm{r0},\omega_\mathrm{i0})$ and the same 
root finding algorithm.
This leads to the subsequent triple of $(\omega_\mathrm{r1},\omega_\mathrm{i1}, k_1)$.
This process continues iteratively until two sufficiently smooth curves,
$(\omega_\mathrm{r},k)$ and $(\omega_\mathrm{i},k)$, are found for a given range of $k$,
which constitute the frequency and growth rate of a specific wave mode.
Numerically, these curves represent one solution of the dispersion relation for each initial intersection $(\omega_\mathrm{r0},\omega_\mathrm{i0}, k_0)$.
This algorithm was validated by reproducing the relativistic dispersion relation curves of \citet{Godfrey1975a}, in particular for arbitrary magnitude of $\omega_\mathrm{i}$ with respect to $\omega_\mathrm{r}$, and some results of \citet{Rafat2019}.

Due to the high precision of the used root finding methods, the uncertainties on the results are very small. This implies that those uncertainties are not visible in the following plots. For the plots showing the maximum growth rates, the uncertainties are enhanced by the limited number of steps in $k$ for which a solution is found.
This implies that the actual maximum might be missed due to a too large step size. However, the combination of both uncertainties is not visible in the corresponding plots either.

A comparison between numerical solutions and results from simulations is possible by considering the maximum values obtained for the respective growth rates. 
In addition, to have a comparable quantity that considers the full range of wave growth in $k$-space, it is possible to define an `integrated growth rate' $\Gamma$: It is defined in such a way that it reveals information about the total electric energy growth:
\begin{equation}
\Gamma=\frac{\int_{z}E\left(z\right)\omega_\mathrm{i}\left(z\right)\mathrm{d}z}{\int_{z}E\left(z\right)\mathrm{d}z}\text{~~.}
\label{eq:integrated}
\end{equation}

The above integral is evaluated for all phase velocities $z$ in the range of positive wave growth. For comparison with simulation results, each growth rate is multiplied by the relativistic kinetic energy corresponding to the respective phase velocity, $E\left(z\right)=\left(\gamma-1\right)m_\mathrm{e}c^2,\, \gamma=\sqrt{1 + z^2}$, and the integral over all kinetic energies is used as a normalisation. Both of the above mentioned quantities, the maximum and integrated growth rates, are considered when comparing the predictions of the linear dispersion theory
with the numerical simulation results (Sect.~\ref{sec:results}).

\subsection{PIC simulations}
\label{subsec:PIC}
We simulated the evolution of streaming instabilities for a wide range of different parameters (beam
Lorentz factor, beam and background temperatures, and beam-to-background density ratio) that
might be characteristic for the plasma of pulsar magnetospheres.
For the fully kinetic simulations of streaming instabilities we utilise the fully kinetic PIC code ACRONYM \citep{Kilian2012}\footnote{http://plasma.nerd2nerd.org/papers.html}. 
ACRONYM contains some features that allow for the elimination of numerical problems and/or instabilities that might occur in kinetic simulations of relativistic
plasmas such as those of pulsar magnetospheres.
In particular, ACRONYM can properly handle highly relativistic velocities of particles and wave propagation in relativistic plasmas.
For our purpose, we use the one-dimensional version of the code.
The code employs the Cole-K\"arkk\"ainen (CK) non-standard finite difference discretisation
scheme (NSFD) for the Maxwell field solver \citep{Cole1997,Karkkainen2006}.
This Maxwell solver minimises the numerical dispersion of
light waves and, thus, any spurious numerical Cherenkov radiation that might occur in
relativistically drifting plasma populations, as it is the case for pulsar plasmas.
Beyond that, our PIC code also utilises a recently developed and specialised shape function to reduce
the numerical Cherenkov radiation, that is to say a higher-order
interpolation scheme known as 
the `Weighting with Time-step dependency' (WT) scheme \citep{Lu2020} in its fourth order version.
To avoid any influence of the choice of shape function in a PIC code on the conservation properties of the system, we choose a charge-conserving current density deposition scheme following \citet{Esikperov2001}.
This approach allows us to use the WT scheme without affecting energy conservation.
We verified that the total energy conservation of this setup is good, while providing an exact conservation of momentum and charge and suppressing the numerical Cherenkov radiation.
\begin{table*}
	\begin{center}
		\def~{\hphantom{0}}
		\caption{Synopsis of the relevant numerical and physical parameters underlying the PIC simulations: Simulation reference number, beam Lorentz factors $\gamma_b$,
		inverse background ($\rho_0$) and beam ($\rho_1$) temperatures,
		simulation length $L$ in number of grid cells ($\Delta$),
		number of particles per cell per species (electrons and positrons)
		of the background $N_0$ and of the beam $N_1$,
		beam-to-background particle density ratio  $r_\mathrm{n}$,
		integrated growth rates from PIC simulations $\Gamma$,
		maximum found growth rate in $k$-space $\omega_\mathrm{i,max,sim}$,
		and the fractional bandwidth $\Delta \omega_\mathrm{r} / \omega_\mathrm{r}$.}.
		\begin{tabular}{crrrrrrlrrr}
            \hline \hline
			Sim. \\ No. & $\gamma_\mathrm{b}$ & $\rho_0$ & $\rho_1$ & L [$\Delta$] & $N_0$ & $N_1$ & \multicolumn{1}{c}{$r_\mathrm{n}$} & \multicolumn{1}{c}{$\Gamma / \omega_\mathrm{p}$} & \multicolumn{1}{c}{$\omega_\mathrm{i,max,sim} / \omega_\mathrm{p}$} & $\Delta \omega_\mathrm{r} / \omega_\mathrm{r}$\\
			\hline
			1 & 26 & 1 & 1 & $10^4$ & $10^4$ & 260 & 0.001 & $(5.1 \pm 0.5) \times 10^{-4}$ & $(8.7 \pm 0.2) \times 10^{-4}$ & $0.12\pm0.02$ \\ 
			2 & 60 & 1 & 1 & $10^4$ & $10^4$ & 600 & 0.001 & $(9.6 \pm 0.6) \times 10^{-4}$ & $(1.28 \pm 0.09) \times 10^{-3}$ & $0.09\pm0.02$ \\ 
			3 & 103 & 1 & 1 & $10^4$ & $10^4$ & 1030 & 0.001 & $(1.07 \pm 0.04) \times 10^{-3}$ & $(1.34 \pm 0.02) \times 10^{-3}$ & $0.09\pm0.01$ \\ 
			4 & 26 & 1 & 1 & $10^5$ & $10^3$ & 1000 & 0.0385 & $(3.68 \pm 0.03) \times 10^{-3}$ & $(5.24 \pm 0.22) \times 10^{-3}$ & $0.38\pm0.06$ \\ 
			5 & 60 & 1 & 1 & $10^5$ & $10^3$ & 1000 & $0.0167$ & $(3.72 \pm 0.09) \times 10^{-3}$ & $(4.61 \pm 0.06) \times 10^{-3}$ & $0.29\pm0.02$ \\ 
			6 & 103 & 1 & 1 & $10^5$ & $10^3$ & 1000 &  $0.00971$ & $(2.80 \pm 0.02) \times 10^{-3}$ & $(3.40 \pm 0.08) \times 10^{-3}$ & $0.23\pm0.04$ \\ 
			7 & 26 & 1 & 10 & $10^	4$ & $10^4$ & 260 & $0.001$ & $(2.76 \pm 0.04) \times 10^{-3}$ & $(3.18 \pm 0.06) \times 10^{-3}$ & $0.17\pm0.03$ \\ 
			8 & 26 & 1 & 100 & $10^4$ & $10^4$ & 260 & $0.001$ & $(3.31 \pm 0.06) \times 10^{-3}$ & $(3.54 \pm 0.09) \times 10^{-3}$ & $0.19\pm0.06$ \\
			9 & 26 & 10 & 10 & $10^4$ & $10^4$ & 260 & $0.001$ & $(7.16 \pm 0.07) \times 10^{-3}$ & $(6.83 \pm 0.05) \times 10^{-3}$ & $0.08\pm0.02$ \\
			10 & 26 & 100 & 100 & $10^4$ & $10^4$ & 260 & $0.001$ & $(7.02 \pm 0.07) \times 10^{-3}$ & $(6.15 \pm 0.04) \times 10^{-3}$ & $0.09\pm0.04$ \\
			11 & 26 & 10 & 1 & $10^4$ & $10^4$ & 260 & $0.001$ & $(4.28 \pm 0.06) \times 10^{-3}$ & $(4.43 \pm 0.06) \times 10^{-3}$ & $0.07\pm0.01$ \\
			12 & 26 & 100 & 1 & $10^4$ & $10^4$ & 260 & $0.001$ & $(5.10 \pm 0.06) \times 10^{-3}$ & $(5.39 \pm 0.06) \times 10^{-3}$ & $0.11\pm0.05$ \\
			13 & 26 & 1 & 1 & $10^4$ & $10^3$ & 2600 & $0.1$ & $(6.06 \pm 0.05) \times 10^{-3}$ & $(7.89 \pm 0.11) \times 10^{-3}$ & $0.65\pm0.14$ \\
			14 & 26 & 1 & 1 & $10^4$ & $10^3$ & 7800 & $0.3$ & $(10.33 \pm 0.09) \times 10^{-3}$ & $(12.37 \pm 0.05) \times 10^{-3}$ & $0.81\pm0.20$ \\
			15 & 26 & 1 & 1 & $10^4$ & $10^3$ & 26000 & $1$ & $(15.79 \pm 0.04) \times 10^{-3}$ & $(17.06 \pm 0.10) \times 10^{-3}$ & $1.16\pm0.18$ \\
			\hline
		\end{tabular}
		\label{tab:simulations}
	\end{center}
\end{table*}

The electron-positron pair plasma is initialised by one-dimensional Maxwell-J\"uttner velocity
distribution functions given by Eq.~\ref{reldisp-mj0} for the background and by Eq.~\ref{reldisp-mj1} for the beam.
Particles are generated using the rejection method in an interval that is larger than the expected velocity range of the distribution function. 
We compared the resulting initial particle distribution
functions with the analytical ones and found a good agreement.
We also tested the convergence using higher numbers of particles, 
finding that the number of particles per cell as given in Table~\ref{tab:simulations}  is sufficient
to accurately describe the initial distribution function.

We conducted 15 simulations to cover the range of the investigated parameters.
Table~\ref{tab:simulations} summarises the main numerical and physical parameters of the simulations.
In our simulations, the time step is $\omega_\mathrm{p} \Delta t = 0.03519$ 
and the (normalised) grid size to the particle skin depth corresponds to $\Delta x \,\omega_\mathrm{p} / c = 0.07108$ (or $d_\mathrm{e}/\Delta x = 14.07$).

The Debye length is estimated as
\begin{equation}
 \lambda_\mathrm{D} = \left( \frac{1}{\lambda_0} + \frac{1}{\lambda_1} \right)^{-\frac{1}{2}},
\end{equation}
where $\lambda_0, \lambda_1$ are the background
and beam Debye lengths.
This definition can be expressed in terms of the relativistic
background and beam `sound' speeds,
$c_\mathrm{s0}$ and $c_\mathrm{s1}$, as \citep{Diver2015}:
\begin{equation}
 \lambda_\mathrm{D} = \left[ \frac{\omega_\mathrm{p}^2}{n_0 + r_\mathrm{n}} \left(n_0\frac{1}{c_\mathrm{s0}^2} + r_\mathrm{n}\frac{1}{c_\mathrm{s1}^2} \right) \right]^{-\frac{1}{2}},
\end{equation}
which, for the Maxwell-J\"uttner distribution, becomes:
\begin{equation}
 c_\mathrm{s0,1}^2 = \frac{c^2}{G_\mathrm{1D}}\frac{\mathrm{d}G_\mathrm{1D}}{\mathrm{d}\rho_{0,1}} \left( \rho_{0,1}\frac{\mathrm{d}G_\mathrm{1D}}{\mathrm{d}\rho_{0,1}} + \frac{1}{\rho_{0,1}}  \right)^{-1},
\end{equation}
with 
\begin{equation}
G_\mathrm{1D}(\rho_{0,1}) = \frac{K_2(\rho_{0,1})}{K_1(\rho_{0,1})},
\end{equation}
where the function $G_\mathrm{1D}$ applies to a one-dimensional velocity
distribution with two degrees of freedom (corresponding to electrons and positrons, respectively) and the adiabatic index $\kappa = 4/2$.
For the three-dimensional case, frequently used in astrophysics,
$G_\mathrm{3D} = K_3(\rho)/K_2(\rho)$
has six degrees of freedom
and an adiabatic index of $\kappa = 8/6$.
This sound speed 
approaches $c_\mathrm{s} \rightarrow c$
for $\rho \rightarrow 0$ (ultra-relativistically hot plasma).
For example, the sound speed is $c_\mathrm{s} = 0.827\,c$ for $\rho=1$, 
while $c_\mathrm{s} = 0.171\,c$ for $\rho=100$.
We note that $c_\mathrm{s} \rightarrow \sqrt{k_\mathrm{B}T/m_\mathrm{e} }$, 
where $k_\mathrm{B}$ is the Boltzmann constant,
for $\rho \rightarrow \infty$, which means that the `sound' speed
becomes the standard thermal speed under non-relativistic conditions.
In our simulations, this normalised Debye length varies
in the range $\lambda_\mathrm{D}/\Delta x = 2.74 - 12.8$
for our limiting cases (maximal and minimal temperatures), 
providing at least (in case of $\rho_0 = \rho_1 = 100$) $2740$ background particles and $712$ beam particles per Debye length.

The one-dimensional simulation box extends along the first (or $x$-) direction, 
that is to say the direction of the pulsar magnetic field. 
Periodic boundary conditions are applied. 

Because of the strong magnetic field
that dominates the pulsar magnetosphere, a one-dimensional domain with a one-dimensional velocity
distribution is sufficient to describe its plasma dynamics. 
This also implies that magnetic effects are suppressed
in our simulation, and all quantities have only one-dimensional profiles.
As the particle current is always along the $x$-direction due to its initial velocity distribution, it creates electric currents
and an electric field solely in this direction.
An electric field only in the $x$-direction does not induce
any non-zero time derivative of the magnetic field, meaning that 
$\partial \textbf{B} / \partial t 
= -\nabla \times \mathbf{E}
= -\nabla \times (E_x,0,0)
= (0,\partial E_x / \partial z, - \partial E_x / \partial y)
= (0,0,0)$,
since the spatial derivatives along the $y$- and $z$-directions
of an arbitrary function are always zero in a one-dimensional simulation.
This implies that the system behaves in the same way as for $\delta B / B \rightarrow 0$
and particles are not influenced by any arbitrary magnetic field 
as its direction coincides with the direction of particle movement (for further details see \citet[Sections 4.3 and 4.4]{Melrose2017}).

The background contains $N_0$ electrons and $N_0$ positrons per cell, while 
the beam is composed of $N_1$ electrons and  $N_1$ positrons per cell.
The background and beam populations are individually charge-neutral in each cell, independent of the respective other population.
We tested the total energy conservation of the code, the generation of wave harmonics, the resolution of the frequency and of the wave vector 
for the analysed relativistic plasma regime.
The results of those tests led us to choose either $10^3$ or $10^4$
background particles per cell per species.
Similarly, we found that for good numerical stability, the number of beam particles per cell has to be $>100$ for the lower values ($\sim 26$) of the beam Lorentz factor $\gamma_\mathrm{b}$.
For $\gamma_b \sim 100$, however, the beam distribution function becomes very broad in the Lorentz factor $\gamma$, such that some particles in the tail of the beam distribution function reach $\gamma > 10^3$.
Hence, we used at least $10^3$ beam particles per cell in this case.
We note that the number of particles in the beam depends on its Lorentz factor $\gamma_\mathrm{b}$ and is thus given as
$N_1 = r_\mathrm{n} \gamma_\mathrm{b} N_0$,
such that, for a density ratio of $r_\mathrm{n} = 1$ and for example $N_0 = 10^3$ as in simulation 15,
the number of beam particles $N_1 = 26\cdot N_0 = 26\times10^3$ does not match, but rather
exceeds the number of background particles in the background reference frame. 
On the other hand, the beam density, and thus the number of beam particles (assuming an equal ratio of physical to numerical particles), 
remains equal in the beam reference frame.

The duration of the simulations depends on the time of the instability onset and its growth rates
that are different for each case. 
The total simulation time $\omega_\mathrm{pe}\Delta T$ thus varies in the range between $2000$ and $15000$.
A typical simulation run consumes 5000 CPU hours. 
The simulation domain size spans either $10^{4}$ or $10^{5}$ grid cells (see $L[\Delta]$ in Table~\ref{tab:simulations})
All our simulations have a spectral resolution that is high enough to describe the wave vector space accurate enough.

The integrated growth rates $\Gamma$ obtained from our simulation,
which shall be compared to the numerical integrated growth rates,
are estimated from the energy of the electric field in the time period from the onset of the exponential growth until the start of saturation.
During this time period, the energy evolution can be fitted by the function $F(t) = E_0 + E_1 \mathrm{e}^{2 \Gamma t}$, where $E_0,E_1,\Gamma$ are arbitrary fit constants. From this fit, the growth rates $\Gamma$ of the total electric energy were calculated.
For a better comparison with linear theory, we also analysed the growth rates as a function of the wavenumber $k$ in order to compare with the analytically obtained maximum growth rates $\omega_\mathrm{i}$.
We transformed the profile of the electric field in the configuration space $\vec{E}(x)$ into the wavenumber space $\vec{E}(k)$ by applying a Fast Fourier Transform (FFT) at each time step. 
Then, we computed the time evolution of the electric field $\vec{E}(k,t)$ for 50 equidistant wavenumbers in the range $k c / \omega_\mathrm{p} \in (0,2.5)$.
Subsequently, we performed an exponential fit of the function $f(t) = e_0 + e_1 \mathrm{e}^{\Gamma t}$ for each of them in the same time window\footnote{We note that the growth rate fit function does not have a factor of two in the exponent because we used the electric field amplitude, not the electric energy. For the same reason, we chose different fit constants $e_0,e_1$.}. 
The maximum growth rate $\omega_\mathrm{i,max,sim}$ is then selected as the largest growth rate that has by found by this procedure.

Two different kinds of uncertainty may occur when estimating $\Gamma$ and $\omega_\mathrm{i,max,sim}$ the way it is outlined above. The first source of uncertainty is induced by the exponential fit itself. This uncertainty can be computed from the diagonal elements of the covariance matrix. The second one is due to the selection of the region where the exponential fit is applied. A too broad fitting region leads to an underestimation of the growth rates.
On the other hand, if it is too narrow, only some local fluctuations of the exponential curve may be considered. 
We tested several fitting regions and analysed the anticipated uncertainties.
For simplicity, we conservatively estimated the total uncertainty by multiplying the one from the fit by a factor of two.
We note that this, as a matter of course, does not include uncertainties due to the numerical simulations themselves.

Another quantity to compare with linear theory is the fractional bandwidth $\Delta \omega_\mathrm{r} / \omega_\mathrm{r}$. This quantity cannot be directly estimated from the functional dependence of the frequency on the wavenumber in the simulations. This is because our simulations do not have enough time steps in the growing phase of the instability to sufficiently cover the frequency space. However, we can estimate the fractional bandwidths from our simulations in a different way:
The fractional bandwidth is first estimated as the full width at half maximum (FWHM) from the growth rates as a function of the wavenumber, $\Delta k$, normalised to the wavenumber of the maximal growth rate $k (\omega_\mathrm{i,max})$, that is to say $\Delta k/ k (\omega_\mathrm{i,max})$.
By assuming that the positive growth rates are present as waves that have a phase speed $v_\phi \approx c$, a linear relation between frequency and wavenumber holds.
This allows us to map the wavenumber range into a frequency range, meaning that
 $\Delta \omega_\mathrm{r} / \omega_\mathrm{r} = \Delta k / k$.
To obtain an estimation of the uncertainties inherent in the calculation of this quantity,
each fractional bandwidth value is calculated for two growth rate profiles
as the FWHM of the functions $\omega_\mathrm{i}(k) \pm \omega_\mathrm{i,err}(k)$,
where $\omega_\mathrm{i,err}(k)$ are the uncertainties of the growth rates estimated
from the fit described above.
The resulting fractional bandwidth is the mean of these two values of the FWHM, while the uncertainty of the fractional bandwidth corresponds to the standard deviation.

The enhanced PIC shot noise (compared to the real thermal fluctuations) due to the limited number of particles per Debye length
in a cold plasma appears not to influence the instability
evolution significantly.
Instead, our results are not very sensitive to the temperature 
for the cold plasma limit $\rho \rightarrow \infty$
(Figs.~\ref{fig:rhobeam}-\ref{fig:rhobackground}).
We tested the convergence of simulations 6 and 10
in order to cover a large range of temperatures, beam velocities,
and numbers of particles per Debye length in this convergence study.
In doing so, we either increased the number of particles by a factor of eight,
or increased the domain size by a factor of ten,
and re-estimated their integrated growth rates, growth rates,
and fractional bandwidths.
All of them are consistent with the values given in Table~\ref{tab:simulations} within the estimated error interval.

\section{Results} 
\label{sec:results}

First of all, we assume an inverse temperature of $\rho_{0}=\rho_{1}=1$ for both beam and background plasma (corresponding to a temperature of $T\approx 10^{10}$~K) and a beam streaming with a Lorentz factor of $\gamma_\mathrm{b} / \gamma_\mathrm{b,min}=2$, where $\gamma_\mathrm{b,min}$ is the minimum value of $\gamma_\mathrm{b}$ at which the background and beam distributions are, according to Penrose's instability criterion of Eq.~\eqref{eq:penrose}, just well enough separated.
The latter criterion, however, is not valid for $\rho_0 < 1$: In this case, the distributions of beam and background plasma show a significant overlap. We note that in all figures of this Section, frequencies are always normalised to the plasma frequency $\omega_\mathrm{p}$ while wavenumbers $k$ are given in units of the electron skin depth $d_\mathrm{e} = c / \omega_\mathrm{p}$.

\begin{figure*}
	\centering
	\includegraphics[width=1.0\textwidth]{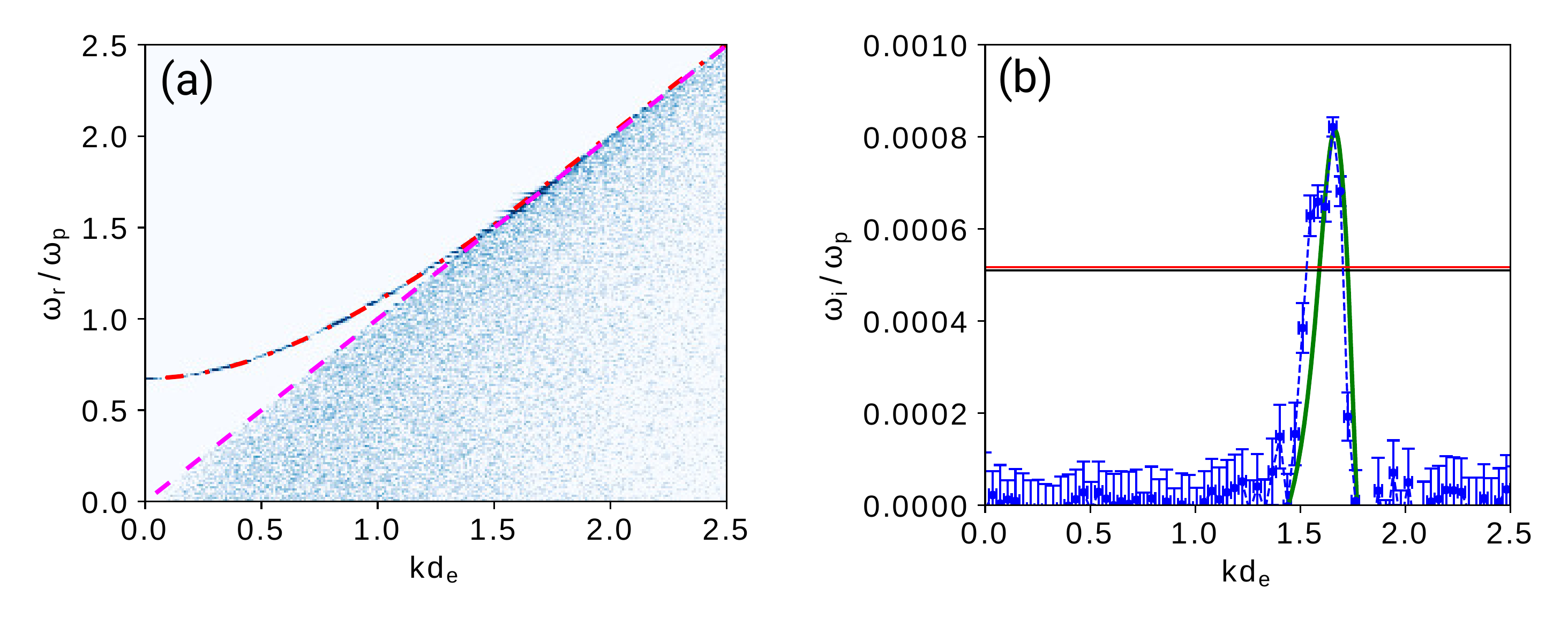}
	\caption{Comparison between a solution obtained by linear theory with the one from simulation for $\rho_{0} = \rho_{1} = 1$, $\gamma_\mathrm{b}=26$, $r_\mathrm{n}=10^{-3}$.\hspace{\textwidth}
	(a) Dispersion diagram showing the real part of the frequency $\omega_\mathrm{r}$ as a function of the wavevector $k$. The background in blue colour scale
	represents the power spectral density obtained from the simulation electric field
    $\vec{E}(x,t)$. The discontinuous lines represent solutions 
	of the linear dispersion relation Eq.~\ref{eq:dispersion}: the
	superluminal L-mode branch (dash-dotted red line) and the subluminal branch (dashed magenta line).\hspace{\textwidth}
	(b) Imaginary part of the frequency $\omega_\mathrm{i}$ (corresponding to the growth rates) as a function of the wavenumber $k$ for the subluminal branch. The green line shows the solution of the linear dispersion relation, and the dashed blue lines with error bars indicate the simulation results.
	The horizontal lines represent the values of the integrated growth rates obtained from the dispersion relation (red line) and from the simulation (black line).
	}
	\label{fig:omega}
\end{figure*}

Fig.~\ref{fig:omega} shows a comparison of a numerical solution of the linear dispersion relation with the results of the PIC Simulation~1, displaying both the real ($\omega_\mathrm{r}$) and imaginary ($\omega_\mathrm{i}$) parts of the frequency as a function of $k$ in the range where a streaming instability is excited.
The real solution includes two dispersion branches:
the superluminal L-mode branch, depicted by a dash-dotted red line, and subluminal branch, represented as a dashed magenta line. 
Beyond that, the figure depicts the FFT $\vec{E}(k,\omega)$ of the power of the electric field oscillations $\vec{E}(x,t)$ obtained by our simulations. It represents the power spectral density (blue colour).
This quantity was obtained by analysing the evolution
of the electric field over 26000 equidistantly selected time steps in the time interval $(0 - 3250)\,\omega_\mathrm{pe} t$.
Thus, Fig.~\ref{fig:omega}a  demonstrates that the regions in the $\omega-k$ diagram with the strongest power spectral density agree 
very well with the locations of the wave modes as predicted by the linear dispersion relation.

The growth rates of both approaches are compared for the subluminal branch in Fig.~\ref{fig:omega}b. This is the wave mode to be further investigated in the following subsections.
The maximum value of the growth rate $\omega_\mathrm{i} / \omega_\mathrm{p} = 8.7 \times 10^{-4}$ occurs at a wavenumber of $k d_\mathrm{e}\approx 1.66$.
The wing of the growth rate curve to the left of its peak value is slightly higher in the simulation than in the linear dispersion relation, while the right wing has almost the same shape in both approaches.
The integrated growth rate computed analytically, shown as a horizontal red line, is in good agreement with the one from simulation (horizontal black line).
In all our cases, the integrated growth rate $\Gamma$ is lower than or equal to the maximum growth rate $\omega_\mathrm{i,max}$.
The former may only reach the maximum growth rate in those cases in which the $\omega_\mathrm{i} - k$ dependence features a very narrow peak.
The parameters used in Fig.~\ref{fig:omega} will be kept constant in the following while separately changing a single parameter in order to assess its effects on instability growth and wave generation.

\subsection{Beam density}
\label{subsec:en}
\begin{figure*}
	\centering
	\includegraphics[width=1.0\textwidth]{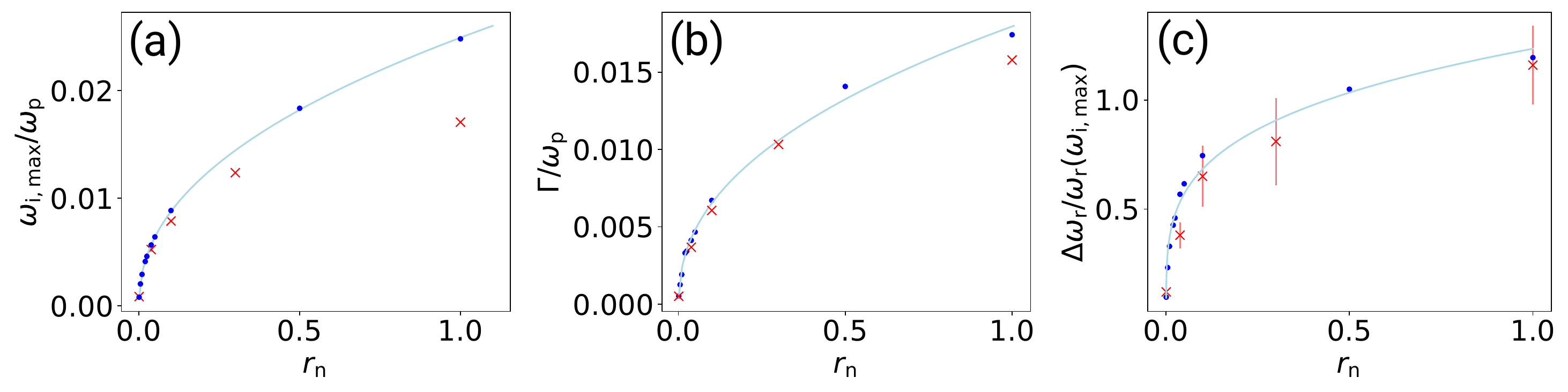}
	\caption{Influence of the beam density on the wave growth and fractional bandwidths, investigated by varying the beam-to-background density ratio $r_\mathrm{n}$ from $10^{-4}$ to $1$:
	(a) Maximum growth rates $\omega_\mathrm{i}$ for different beam-to-background density ratios $r_\mathrm{n}$ obtained by solving the linear dispersion relation (dots) and from simulations (crosses) with fit function (solid line).
	(b) Integrated growth rates calculated from numerical solutions of the dispersion relation (dots) and from simulation results (crosses) with fit function (solid line).
	(c) Fractional bandwidth $\Delta \omega_\mathrm{r}$ at the point where $\omega_\mathrm{i}$ is at its maximum, normalised to the corresponding real part of the frequency $\omega_\mathrm{r}$, obtained by linear theory (dots) and simulation (crosses). 
	All three figures are for $\rho_{0}=\rho_{1}=1$ and $\gamma_\mathrm{b}=26$. Details regarding the fit functions can be found in Table~\ref{tab:simulations}.}
	\label{fig:en}
\end{figure*}
To begin with, the question of the dependence of the wave growth on the beam-to-background density ratio is addressed.
We vary the parameter $r_\mathrm{n}$ from very small values ($r_\mathrm{n}=10^{-4}$, a weak beam) up to the scenario of a strong beam ($r_\mathrm{n}=1$), while keeping the beam drift speed constant at $\gamma_\mathrm{b} = 26$.

Fig.~\ref{fig:en} illustrates the effect of altering the density ratio on the wave growth, showing the maximum as well as integrated growth rates and fractional bandwidths as a function of different density ratios $r_\mathrm{n}$. 
We note that for $r_\mathrm{n} = 1$, the beam plasma has to be represented by 26000 (macro-) particles per cell, which means it contains 26 times more particles than the background.
The maximum and integrated growth rates as well as the fractional bandwidths grow with increasing beam density, but not linearly.
The agreement between simulation and analytical theory proves to be better in case of the integrated growth rates that follow the analytical trend.
While the simulation maximum growth rates are very similar to the linear dispersion relation calculations for small $r_\mathrm{n}$, they tend to be lower than the analytical calculations for higher density ratios.
The fractional bandwidths from simulations are in reasonable agreement with those from linear theory with a slight trend towards lower values than the analytical ones.

To predict the maximum growth rates that may occur for a certain density ratio, a fit function of the following form suits the observed dependence:
\begin{equation}
\label{eq:en}
f\left(r_\mathrm{n}\right)=a\cdot\left(r_\mathrm{n}+b\right)^{d},
\end{equation}
where $a$, $b$, and $d$ are constants.
If the growth rate is zero for a case without beam, the parameter $b$ should be zero. In our case, it approaches zero.
The growth rate has a lower than linear dependence on the beam-to-background density ratio, 
which is reflected by a fit parameter $d < 1$.
The integrated growth rates and bandwidths follow the same behaviour as the maximum growth rates, such that they can be described a fit function of the same shape with different values for the parameters $a$, $b$ and $d$. 
A summary of these fit parameters is to be found in Table~\ref{tab:rhobeam}.

\subsection{Beam temperature}
\label{subsec:rhobeam}

This subsection is dedicated to the results with regard to the influence of the (inverse) beam temperature $\rho_1$
on the growth rates $\omega_\mathrm{i;max}$ and $\Gamma$ as well as on the fractional bandwidths $\Delta\omega_\mathrm{r}$ at the
frequency of maximum growth.

\begin{figure*}
	\centering
	\includegraphics[width=1.0\textwidth]{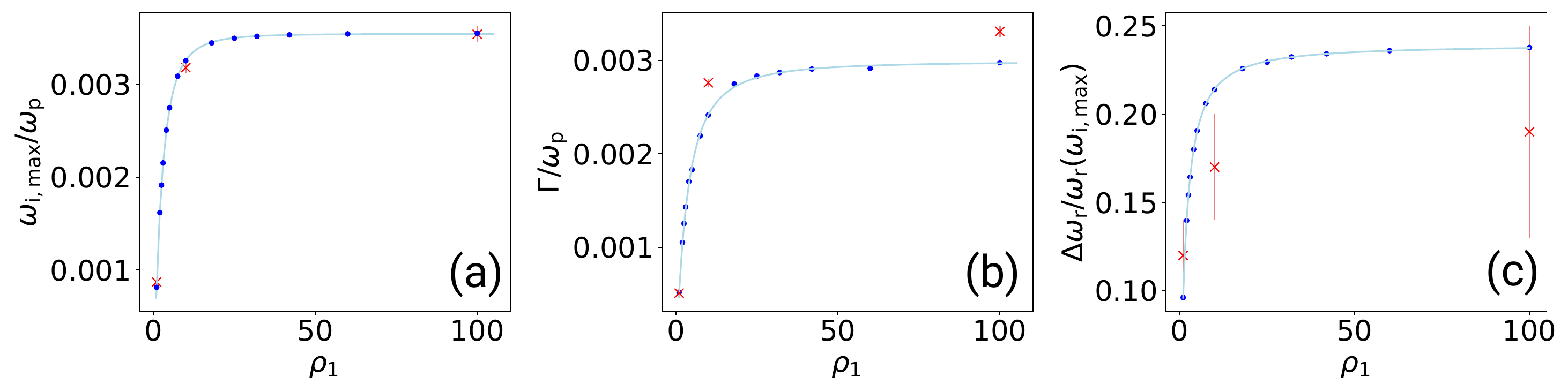}
	\caption{Influence of the (inverse) beam temperature $\rho_1$: (a) Maximum growth rates $\omega_\mathrm{i}$ for different beam inverse temperatures $\rho_{1}$ obtained from the analytical solution of the linear dispersion relation (dots) and from simulations (crosses) with fit function (solid line).
	(b) Integrated growth rates from solutions of the dispersion relation (dots) and from simulations (crosses) with fit function (solid line).
	(c) Fractional bandwidth $\Delta \omega_\mathrm{r}$ at the point where $\omega_\mathrm{i}$ is at its maximum, normalised to the corresponding real part of the frequency $\omega_\mathrm{r}$ obtained by linear theory (dots) and simulation (crosses).
	All three figures are for $\rho_{0}=1$,$\gamma_\mathrm{b}=26$ and $r_\mathrm{n}=10^{-3}$. All fit parameters are included in Table~\ref{tab:simulations}.}
	\label{fig:rhobeam}
\end{figure*}
Fig.~\ref{fig:rhobeam} shows the behaviour of these growth rates and bandwidths depending on the beam temperature, varying it between $\rho_1=1$ and $\rho_1=100$ in both analytical theory and simulation.
For beam temperatures approaching the non-relativistic regime, the maximum growth rate firstly increases until it approaches a saturation level for the limit of a very cold beam, $\rho_1 \rightarrow \infty$.
A similar dependence can be found for the integrated growth rates and fractional bandwidths.
A fit function of the following form describes this behaviour:
\begin{equation}
\label{eq:rhobeam}
f\left(\rho_{1}\right)=a-\frac{b}{\left(c+\rho_{1}\right)^d}\text{,}
\end{equation}
where the parameter $a$ corresponds to the saturation value that can be reached for the beam temperature approaching zero. This saturation occurs at $\rho_{1}\approx18$.
The parameters $b$ and $d$ determine how quickly the growth rate converges to this saturation value with decreasing temperature. 
This way, Eq.~\eqref{eq:rhobeam} quantifies the increase in growth of the instability with higher $\rho_1$ due to the increasing narrowness and steepness of the beam velocity distribution function. The specific values of the fit parameters are indicated in Table~\ref{tab:rhobeam}.

Comparing the analytical results with our simulations, it can be stated that the maximum growth rates from simulation agree with the solutions of the linear dispersion relation very well, while the simulation integrated growth rates slightly exceed the analytical ones.
In the case of the fractional bandwidths, the simulation values are below those obtained by linear theory. However, they have large uncertainties that include the linear theory results in two of the three cases.

\subsection{Background temperature}
\label{subsec:constrho1}

\begin{figure*}
	\centering
	\includegraphics[width=1.0\textwidth]{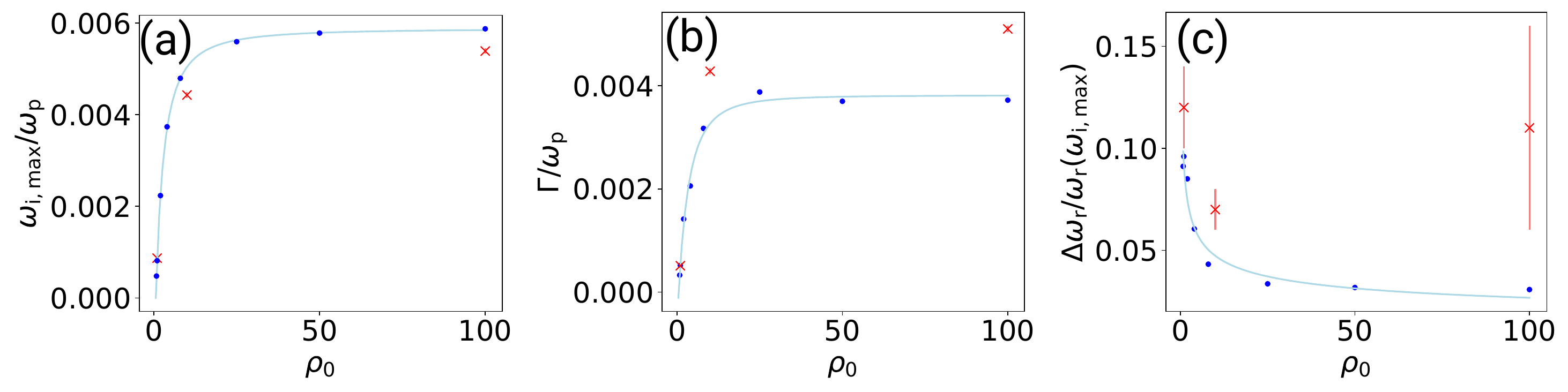}
	\caption{Influence of the (inverse) background temperature $\rho_{0}$; same as Fig.~\ref{fig:rhobeam}, but varying $\rho_{0}$ instead of $\rho_{1}$ for $\rho_{1}=\mathrm{const.}=1$.}
	\label{fig:constrho1}
\end{figure*}

A similar investigation can be carried out by changing the temperature of the background plasma while keeping the beam temperature constant at $\rho_{1}=1$. The obtained maximum and integrated growth rates also obey the same behaviour as those in Sect.~\ref{subsec:rhobeam} and can thus be described by a fit function of the same form Eq.~\ref{eq:rhobeam}. These fit functions are displayed in Fig.~\ref{fig:constrho1}, the parameters being stated in Table~\ref{tab:rhobeam}.  
Also in this case, saturation is observed at $\rho_{0}\approx18$. However, a different behaviour is found for the fractional bandwidths: They decrease with lower background temperatures. 
With respect to the agreement between analytical theory and simulation, the same tendency as in Sect.~\ref{subsec:rhobeam} is observed in the case of the integrated growth rates, namely that the simulation results are larger than the linear theory values. Meanwhile the agreement is better for the maximum growth rates, with the simulation growth rates however being slightly lower than the theoretical predictions.
In the case of the fractional bandwidths, however, the simulation results differ significantly from those obtained by linear theory. They do not decrease as much as predicted by the theoretical curve, more specifically, the simulation results are always higher. In addition, they actually follow the opposite trend as the analytical curve for a very cold background, meaning that the simulation results increase while the theoretical curve decreases.

\subsection{Background and beam temperature}
\label{subsec:rhobackground}
\begin{figure*}
	\centering
	\includegraphics[width=1.0\textwidth]{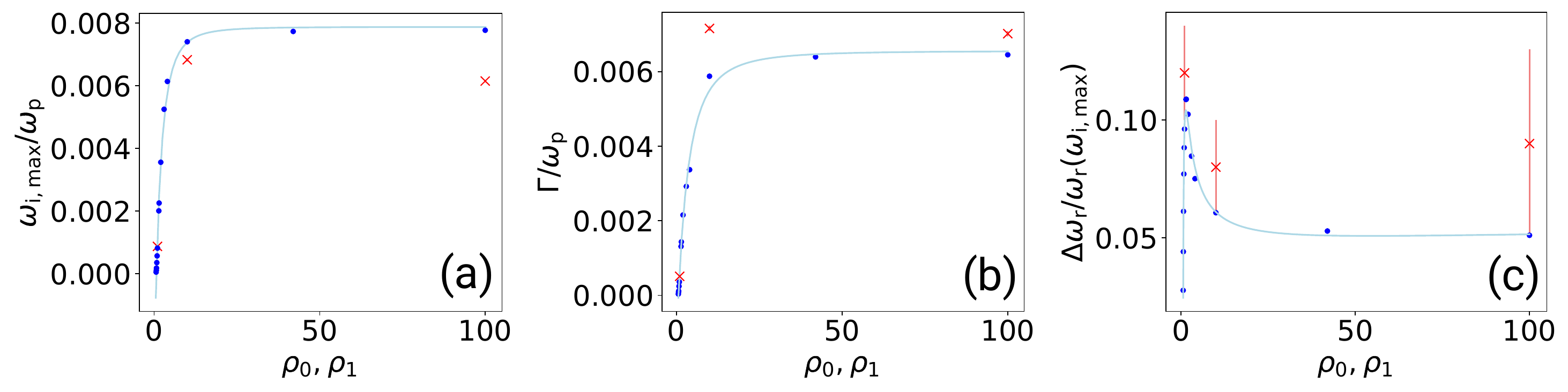}
	\caption{Influence of the (inverse) overall (beam and background) temperature ($\rho_{0} = \rho_{1}$, corresponding to $r_\rho = 1$); same as Fig.~\ref{fig:rhobeam}, but simultaneously changing $\rho_{0}$ and $\rho_{1}$ instead of solely $\rho_{1}$. }	
	\label{fig:rhobackground}
\end{figure*}
Since either the temperature of the beam or the background plasma is always kept at $\rho_{0/1}=1$ in the preceding Sects.~\ref{subsec:rhobeam} and \ref{subsec:constrho1}, it is worth to explore the behaviour for a different overall temperature.
This means that we simultaneously change $\rho_{0}$ and $\rho_{1}$ while their ratio is set to 
$r_\rho=1$.
This temperature ratio corresponds to the one considered in the work by \citet{Rafat2019}. 
The maximum and integrated growth rates as well as fractional bandwidths as a function of the overall temperature are shown in Fig.~\ref{fig:rhobackground}.

A similar explanation as in Sect.~\ref{subsec:rhobeam} applies to the increase in growth rates with decreasing temperature, that is to say it is due to the steepening of both beam and background velocity distribution functions. 
In this case, the growth rates are higher and the rise is steeper than in the cases where the temperature of only one component is decreased.
A saturation of the maximum and integrated growth rates from the solutions of the linear dispersion relation occurs at approximately the same temperature of the beam or background, respectively, as observed in the previously investigated cases. 
However, the maximum and integrated growth rates from simulations - which have again slightly lower maximum and slightly higher integrated values - do not saturate after this value but rather start to decrease again.

The fractional bandwidth shows a different behaviour. After a steep rise until $\rho_{0}\approx1.5$, it starts to decrease again. For values of $\rho_{0} < 1.5$ the behaviour is governed by the beam temperature, for which the bandwidth rises when the beam gets colder, while the background temperature starts to dominate the behaviour for $\rho_{0} > 1.5$. This decrease is also loosely observed in the simulation results. Although their values are always above the analytical ones and diverge significantly for the coldest temperature case, the error bars of the simulation results barely include the analytical curve. In this sense, the disagreement is not as severe as for Figure.~\ref{fig:constrho1}. The different behaviour of the fractional bandwidths leads to the fact that the fit function \eqref{eq:rhobeam} is not sufficient to capture the dependence and has to be extended in the following way:
\begin{equation}
	\label{eq:rhobackground}
	f\left(\rho_{0/1}\right)=a-\frac{1}{\left(\rho_{0/1}+b\right)^c}+\frac{d}{\left(\rho_{0/1}\right)^e}\text{.}
\end{equation}
The specific values of the fit parameters for the curves depicted in Fig.~\ref{fig:rhobackground} are given in Table~\ref{tab:rhobeam}.

\subsection{Beam drift speed for constant $r_\mathrm{n}$}
\label{subsec:gammab}
In the following, we analyse the effects of the beam drift speed on the growth rates.
We note that changing the beam drift speed under the assumption of a weak beam with $r_\mathrm{n}=10^{-3}$ necessarily implies a change of the particle density ratio $n_1 / n_2$
because of the dependence of $r_\mathrm{n}$ on 
the beam Lorentz factor.

\begin{figure*}
	\centering
	\includegraphics[width=1.0\textwidth]{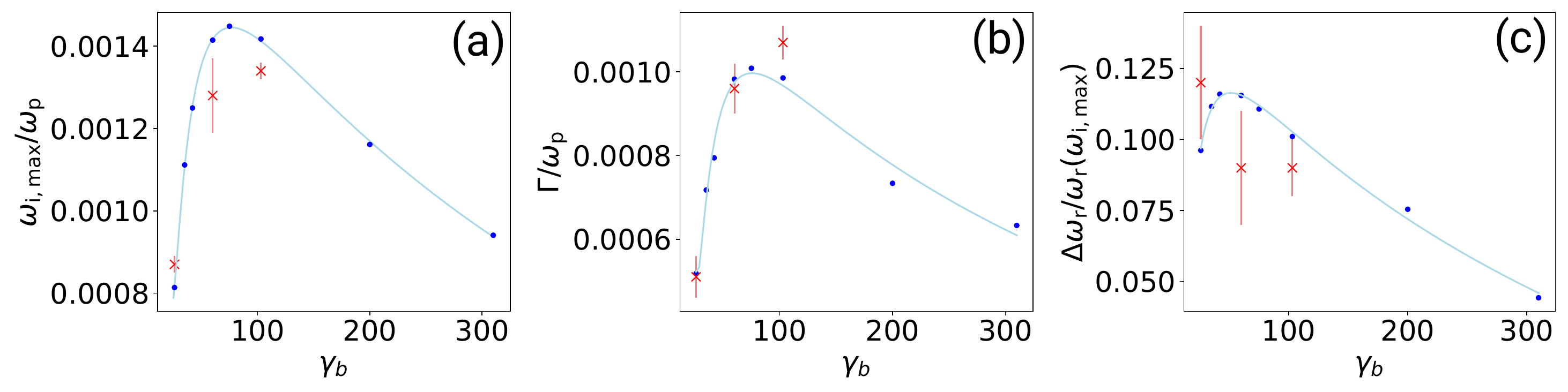}
	\caption{Influence of the beam velocity in the weak-beam limit $n_1 \ll n_0$ (i.e. $r_\mathrm{n} = 10^{-3}$). (a) Maximum values of $\omega_\mathrm{i}$ for different beam velocities $\gamma_\mathrm{b} = (1 - \beta_\mathrm{b}^2)^{-\frac{1}{2}}$ obtained from the solutions of the linear dispersion relation (dots) and simulations (crosses) with fit function (solid line).
	(b) Integrated growth rates from numerical solutions of the dispersion relation (dots) and from simulations (crosses) with fit function (solid line).
	(c) Fractional bandwidth $\Delta \omega_\mathrm{r}$ at the point where $\omega_\mathrm{i}$ is at its maximum, normalised to the corresponding real part of frequency $\omega_\mathrm{r}$, obtained from linear theory (dots) and simulation (crosses).
	All three figures for $\rho_{0}= \rho_1 = 1$, $r_\mathrm{n}=10^{-3}$.
	}
	\label{fig:gammab}
\end{figure*}

As it can be inferred from Fig.~\ref{fig:gammab}, firstly, the growth rates, the integrated growth rates as well as the fractional bandwidths increase with faster beam velocities until they reach a maximum value for $\gamma_\mathrm{b} \approx 75$.
Subsequently, they decrease with increasing beam velocity.
The extension of the fit function \eqref{eq:rhobeam}, given by Eq.~\ref{eq:rhobackground}, can also be applied here to describe the dependence of the growth rates as well as bandwidths on $\gamma_\mathrm{b}$.
The second term of Eq.~\eqref{eq:rhobackground} is connected with the increase in growth rates for $\gamma_\mathrm{b} < 75$ while the third term describes the growth rate decrease for $\gamma_\mathrm{b} > 75$.
The corresponding parameters of the fitted curves are given in Table~\ref{tab:rhobeam}.
In the case of the simulational growth rates, it cannot be clearly stated whether they follow the exact same behaviour as the analytical ones since they have not, so far, reached the regime where they would decrease again. A reasonable agreement, however, is found for the integrated growth rates at lower $\gamma_\mathrm{b}$, while the value of $\Gamma$ is higher than the analytical solution for a higher $\gamma_\mathrm{b}$. The maximum growth rates follow the trend that is observed for the change of all investigated parameters as they tend to be lower than the analytically calculated ones. Equally, the behaviour of the fractional bandwidths prevents a clear conclusion towards the agreement with linear theory as it appears to rather be opposed to the analytical curve.

\subsection{Beam drift speed for $n_0 = n_1$}
\label{subsec:constdens}
\begin{figure*}
	\centering
	\includegraphics[width=1.0\textwidth]{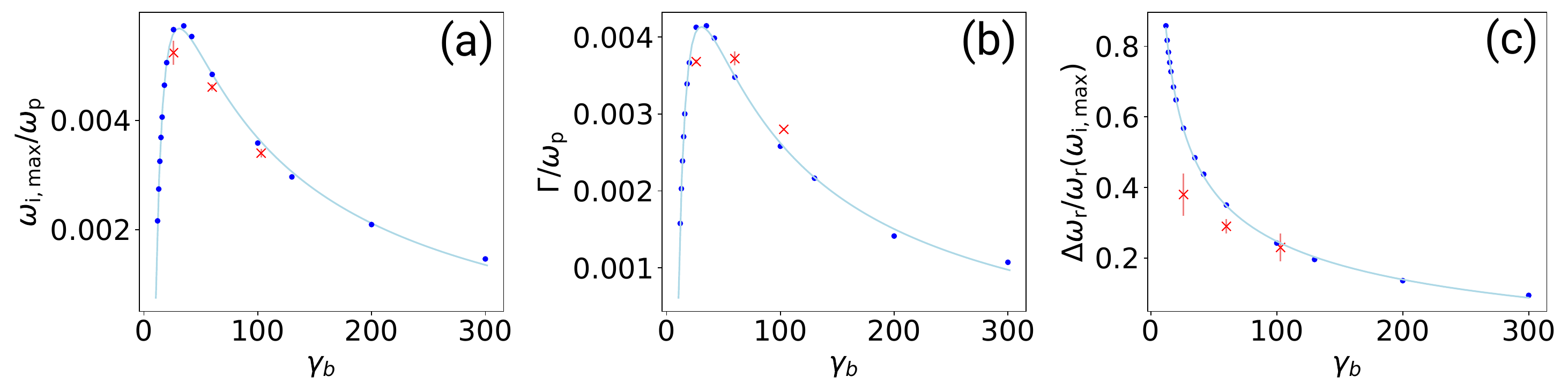}
	\caption{Influence of the beam velocity $\gamma_\mathrm{b}$ in the case of equal beam and background densities ($n_1 = n_0$); Same as Fig.~\ref{fig:gammab}, but for a density ratio of $n_0 = n_1$.}
	\label{fig:constdens}
\end{figure*}

Within the \cite{Usov1987} model, it is sensible to assume approximately equal densities of the beam and background plasma.
Fig.~\ref{fig:constdens} shows the maximum and integrated growth rates and the fractional bandwidths for increasing beam velocities while keeping the ratio of particle number densities $n_0$ and $n_1$ constant. 

As in the case analysed in the preceding Sect.~\ref{subsec:gammab}, the growth rates increase with higher beam velocity, this time up to a value of $\gamma_\mathrm{b} \approx 30$. After reaching the maximum, the growth rates subsequently decrease. Compared to Sect.~\ref{subsec:gammab}, however, this decrease is at first steeper with increasing $\gamma_\mathrm{b}$ until it levels off at large $\gamma_\mathrm{b}$.
The extended fit function \eqref{eq:rhobackground} is suited to describe the behaviour of the growth rates here as well,
assuming the fit parameters summarised in the Table~\ref{tab:rhobeam}. 
By contrast, the fractional bandwidths exhibit a different behaviour than in the case analysed before:
They monotonously decrease for the entire beam velocity range. They can be described by the function \eqref{eq:rhobeam} with fit parameters given in Table~\ref{tab:rhobeam}.
The simulational and analytical growth rates match well for this case. Furthermore, the fractional bandwidths from simulations follow the same trend as those obtained by linear theory, although the curve indicated by the simulation points might be slightly flatter than the analytical one.

\begin{table*}
    \begin{center}
        \def~{\hphantom{0}}
        \caption{
		Parameters of the fitted functions for the growth rates, the integrated growth rates, and the fractional bandwidths.
		}
        \begin{tabular}{ll|rrrrr}
        \hline \hline
		Figure & Equation &$a$  & $b$ & $c$  &   $d$&$e$  \\[3pt]
		\hline
		\multicolumn{7}{l}{\textit{Growth rates}} \\
		\ref{fig:en}(a)& Eq.~\ref{eq:en} & $(2.491\pm0.009)\times10^{-2}$  & $(-6\pm1)\times10^{-4}$ & -- & $(4.51\pm0.03)\times10^{-1}$ &--\\
		\ref{fig:rhobeam}(a)& Eq.~\ref{eq:rhobeam}&$(3.544\pm0.004)\times10^{-3}$  & $1.0\pm0.4$ &$6.6\pm0.4$ & $2.9\pm0.1$ &--\\
		\ref{fig:constrho1}(a)& Eq.~\ref{eq:rhobeam} &$(5.88\pm0.37)\times10^{-3}$ &$(5\pm3)\times 10^{-2}$&$3.3\pm0.6$&$1.6\pm0.2$&--\\
		\ref{fig:rhobackground}(a)& Eq.~\ref{eq:rhobeam} &$(7.9\pm0.2)\times10^{-3}$ &$1.0\pm0.0$&$4.9\pm0.3$&$2.81\pm0.08$&--\\
		\ref{fig:gammab}(a)& Eq.~\ref{eq:rhobackground} & $-19\pm1$ & $(-6.7\pm0.6)\times10^{-4}$ & $(-1.8\pm0.6)\times10^{-1}$& $1.3\pm0.1$ &$1.0049\pm0.0004$\\
		\ref{fig:constdens}(a)& Eq.~\ref{eq:rhobackground}&$2.1\pm0.4$&$1\pm6$&$0.3\pm7$&$0.7\pm4$&$(-2\pm6)\times 10^{-3}$\\
		\hline
		\multicolumn{7}{l}{\textit{Integrated growth rates}} \\
		\ref{fig:en}(b)& Eq.~\ref{eq:en} & $(1.80\pm0.04)\times10^{-2}$ & $(-8\pm5)\times 10^{-4}$ & -- & $(4.4\pm0.2)\times10^{-1}$ &--\\
		\ref{fig:rhobeam}(b)& Eq.~\ref{eq:rhobeam}&$(2.99\pm0.03)\times10^{-3}$&$(9\pm8)\times10^{-2}$&$6\pm1$&$1.8\pm0.3$& --\\
		\ref{fig:constrho1}(a)& Eq.~\ref{eq:rhobeam} & $(3.81\pm0.9)\times 10^{-3}$ & $1\pm0$ & $8.1\pm0.7$ & $2.6\pm0.1$ & --\\ 
		\ref{fig:rhobackground}(b)& Eq.~\ref{eq:rhobeam} & $(6.6\pm0.1)\times10^{-3}$&$1.0\pm0.0$&$7.6\pm0.4$&$2.39\pm0.05$& --\\
		\ref{fig:gammab}(b)& Eq.~\ref{eq:rhobackground}&$-23\pm3$&$(-4\pm2)\times10^{-4}$&$(-4\pm8)\times10^{-1}$&$(1.6\pm0.6)\times10^{-1}$&$1.003\pm0.001$\\
		\ref{fig:constdens}(b)& Eq.~\ref{eq:rhobackground}& $2\pm71$ & $1 \pm 10$ & $(3\pm1)\times10^{-1}$ & $(8 \pm 6)\times 10^{-1}$ & $(-7\pm0.4)\times10^{-4} $\\
		\hline
		
		\multicolumn{7}{l}{\textit{Fractional bandwidths}} \\
		\ref{fig:en}(c) & Eq.~\ref{eq:rhobeam} &$1.24\pm0.04$&  $(-9.6\pm0.0)\times 10^{-4}$&--&$(2.6\pm0.1)\times10^{-1}$&--\\
		\ref{fig:rhobeam}(c) & Eq.~\ref{eq:rhobeam} & $0.2392\pm0.0003$ & $0.51 \pm 0.03$ & $1.83\pm0.09$ & $1.22\pm 0.03$&--\\
		\ref{fig:constrho1}(c) & Eq.~\ref{eq:rhobeam} & $0.08\pm0.02$ & $1 \pm 0$ & $0.3\pm0.2$ & $0.009\pm 0.022$&--\\
		\ref{fig:rhobackground}(c) & Eq.~\ref{eq:rhobackground} & $-0.2\pm0.1$ & $0.5\pm0.3$ & $1.14\pm0.34$ & $0.52\pm0.06$ & $0.07\pm0.06$ \\
		\ref{fig:gammab}(c) & Eq.~\ref{eq:rhobackground} & $(-1.6\pm0.4)\times10^{-2}$ & $(-6\pm1)\times 10^{-2}$ & $(-3 \pm 2)$ & $(9\pm7)\times 10^{-1}$ & $1.4\pm0.1$ \\
		\ref{fig:constdens}(c)& Eq.~\ref{eq:rhobeam}&$-0.23\pm0.06$&$-2.6\pm0.3$&$-1\pm1.3$&$0.37\pm0.05$&--\\
		\hline
		\end{tabular}
        \label{tab:rhobeam}
    \end{center}
\end{table*}

\section{Discussion}
\label{sec:discussion}
To study one of the possible radio emission mechanisms of pulsars, we used a model based on the
mutual interaction of two relativistic beams \citep{Usov1987}. 
We investigated the properties of this kinetic instability for a wide range of plasma parameters,
using an appropriate one-dimensional Maxwell-J\"uttner velocity distribution.
As an initial validation test, we compared our linear dispersion results with those by \citet{Rafat2019b} and with PIC simulations results (see Fig.~\ref{fig:omega} as an example). We found only some small discrepancies
between both approaches in the whole investigated parameter range,
mostly caused by non-linear effects present in PIC simulations. Beyond the analytical results, based on the linearisation of the Vlasov equation, our simulations also include the saturation of the instability after the linear growing phase, when the exponential wave growth becomes too large for the linear approximation to be valid. Once the wave amplitude stops to grow, the system is dominated by those non-linear effects that might be responsible for the observed differences. With both approaches, we obtained positive growth rates for the entire investigated parameter range.

Now the question arises whether this instability can account for the observed strong pulsar radio emission.
According to \citet{Ursov1988}, the time for instability development must be at least one order smaller than the typical lifetime of the system --- the time span between two plasma bunch releases. 
This time span until the next bunch release, $\Delta t$, can be estimated to be of the order of ${10}\,\mu$s \citep{Ruderman1975,Asseo1998}. 
Thus, in order to quantify the effectiveness of the instability, it is possible to estimate the growth time of the instability assuming a growth factor of $G=1$ (Eq.~\eqref{eq:def_g}), which means at the transition between efficient and inefficient wave growth.
For this, we compute $\omega_\mathrm{i}$ for this case by taking the inverse of the imaginary part of the obtained frequency. 
If $\Delta t = G / \omega_\mathrm{i} < {10}\ \mu$s, the instability can be considered effective.
The plasma frequency can be estimated as $\omega_\mathrm{p} \approx 3.6 \times 10^{9}\,s^{-1}$ \citep{Ursov1988} at the distance from the pulsar where the instability is expected to develop.
We assume that the normalised growth rate value $\omega_\mathrm{i} / \omega_\mathrm{p}$ is invariant under Lorentz transformations.
Then the time interval $\Delta t = 10\,\mu$s determines a threshold for the minimum growth rate in the pulsar frame: $\omega_\mathrm{i,thr}/\omega_\mathrm{p} = 1 / (\Delta t \, \omega_\mathrm{p}) = 2.8 \times 10^{-4}$ .
Thus, according to this criterion, all unstable waves with growth rates $\omega_\mathrm{i} / \omega_\mathrm{p} > \omega_\mathrm{i,thr}/\omega_\mathrm{p}$ would have sufficient time to develop during the mutual bunch interactions.

We note, however, that this threshold growth rate depends on the underlying model.  For example, \citet{Weatherall1994} states higher values for the typical pulsar plasma density of $n_\mathrm{e}\approx9.7\cdot10^{13}\,$cm$^{-3}$. 
The corresponding plasma frequency $\omega_\mathrm{p}\approx7.9\cdot10^{11}s^{-1}$ would lead to a smaller threshold growth rate of $\omega_\mathrm{i,thr}/\omega_\mathrm{p} = 1.3\times 10^{-6}$, providing an even broader parameter range for a sufficiently strong instability. However, the model by \citet{Weatherall1994} assumes a strong-beam model with counter-streaming electron and positron populations as a major source of the instability that causes the coherent pulsar radio emission. Thus, the threshold given in the preceding paragraph appears to be better applicable to our investigated case and presupposed model.

Comparing this threshold value of $\omega_\mathrm{i,thr,pulsar} / \omega_\mathrm{p} = 2.8\times 10^{-4}$ with our analytical and simulation results for all analysed parameters, the instability appears to be strong enough to develop -- and may thus lead to the generation of significant radio emission -- for density ratios $r_\mathrm{n} > 10^{-3}$ and for the beam and background inverse temperatures $\rho_0,\rho_1 \geq 1$.
For small $\rho_0,\rho_1 < 1$, however, the distribution functions of of both beam and background overlap significantly. 
This temperature range is in agreement with \citet{Weatherall1994}.
Beyond that, the growth rates exceed the threshold for beam velocity Lorentz factors $13 < \gamma_\mathrm{b} < 300$. The beam velocity must be large enough to separate both distributions while the temperature has to be low enough in order to guarantee that the relativistic broadening of the distribution does not quench the instability.
We note that all our simulations have a growth rate that is high enough to overcome the growth rate threshold.

Contextualising the findings of our parametric study within the existing models, we can bridge the gap between those models assuming a weak beam and those focusing on a strong-beam instability.
We show how the growth rates and the fractional bandwidths increase with a higher beam density.
This implies that the velocity distribution slope increases, such that the instability becomes stronger. 
This dependence is not linear.
Since for larger $r_\mathrm{n}$, a deviation between the analytical and simulation growth rate is observed, the extent of the increase in the achievable growth rates with larger beam density is not clear. Nevertheless, we can conclude that an increase in $r_\mathrm{n}$ to $\sim 0.3$ -- which is the point where both simulation and analytical growth rates still agree with each other -- significantly enhances the instability growth rates. 
While the model by \citet{Rafat2019b} explicitly assumes the weak-beam condition, $r_\mathrm{n} \ll 1$,
the theory discussed in \citet{Usov1987} is not limited to just a weak beam. A rather strong-beam model would in principle be conceivable within the \citet{Usov1987} model since both successively emitted bunches initially feature equal densities, and the relative velocity between the plasma that forms the beam and the one that can be considered the background might be not as large as initially suggested by \citet{Usov1987}.

We also study the transition between hot and cold beam and  background distributions.
With decreasing temperature, a colder distribution leads to a larger slope of the corresponding velocity distribution function, causing a stronger instability.
As analysed in Sect.~\ref{subsec:rhobeam}, for low values of the inverse beam temperature ($\rho_1 \approx 1$), the distribution function is very broad and the growth rates are dominated by relativistic effects.
With increasing $\rho_1$, the growth rates approach a saturation  ($\rho_1 > 18$), even though the distribution still significantly deviates from a non-relativistic Maxwellian distribution.
The saturation values are consistent with the behaviour of the growth rates in reactive fluid instabilities within the cold beam approximation \citep{Hinata1976,Shalaby2017}:
\begin{equation}
\frac{\omega_\mathrm{i}}{\omega_\mathrm{p}} = \frac{1 }{2 \sqrt{2 \gamma_\mathrm{b}^3}}~.
\end{equation}
For $\gamma_\mathrm{b} = 26$, this expression yields $\omega_\mathrm{i}/\omega_\mathrm{p} = 2.67\times 10^{-3}$.
From the fit (Table~\ref{tab:rhobeam}; parameter $a$), we can estimate the saturation of our analytically calculated maximum growth rates at $\omega_\mathrm{i}/\omega_\mathrm{p} = (3.542 \pm 0.004 )\times 10^{-3}$ (for $\rho_1 \rightarrow \infty$). 
The simulations for $\rho_1 = 100$ also lead to growth rates near this saturation value ($\omega_\mathrm{i}/\omega_\mathrm{p} =(3.54 \pm 0.08 )\times 10^{-3}$).
The integrated growth rates experience a saturation at $\Gamma/\omega_\mathrm{p} = (2.98 \pm 0.04)\times 10^{-3}$ in the linear calculations and $\Gamma/\omega_\mathrm{p} = (3.31 \pm 0.06)\times 10^{-3}$ in the simulations.
In the case of lowering the background temperature, the saturation value is higher than in the case of decreasing only the beam temperature. 
Due to the plasma bunch interaction, the behaviour of the growth rates when simultaneously changing the temperature of the background and the beam is expected.
The resulting growth rates cannot be easily estimated by a simple interpolation from the growth rate dependence on either the beam or background temperature.
The growth rate saturation value when varying both temperatures is higher than in the case of solely changing one temperature.
Therefore, we expect that for the generation of strong emission, 
a colder temperature of both beam and background than predicted by the current theoretical models could occur, for example, if both interacting bunches have already cooled down when moving away from the pulsar. \\

Moreover, we analyse the growth rates as a function of the beam velocity for two cases. 
The first case is for constant $r_\mathrm{n}=\frac{n_{1}}{\gamma_\mathrm{b}n_{0}}$,
while the second one is for a constant particle density ratio $n_0 / n_1 = 1$.
Throughout the investigated
parameter range, the latter case always exhibits higher growth rates compared to the first one due to the higher beam density in the background reference frame.
In both cases, the growth rates first increase with increasing beam velocity while the distributions are overlapping, until the velocity is sufficiently high for the velocity distributions to fully separate.
The maximum growth rate as a function of the beam velocity is reached for lower beam velocities in the case of $n_0 = n_1$ than for the case of $r_\mathrm{n} = 10^{-3}$.
This is because for the former case, $r_\mathrm{n}$ decreases with increasing beam velocity, thus lowering the growth rates.
In the case $r_\mathrm{n} = 10^{-3}$, the density ratio in the background frame remains constant, and the instability growth rates are not significantly quenched for higher beam velocities due to relativistic effects, that is to say due to a strong broadening of the beam distribution with increasing beam velocity.
The case $n_1 = n_2$, by contrast, represents a closed plasma system where the number of particles of both populations remains constant and only the beam velocity 
varies.
Given the estimated range of Lorentz factors $\gamma$ of the respective bunches as stated in the introduction ($\gamma_\mathrm{min}\sim 10$ to $\gamma_\mathrm{max}\sim 10^{3}-10^{4}$, see Sect.\ref{sec:intro}), our results suggest that the interaction responsible for the pulsar emission may not take place between the fastest particles from the successive bunch ($\gamma_\mathrm{max}$) and the slowest ones from the one emitted before ($\gamma_\mathrm{min}$), as initially suggested in the work of \citet{Usov1987}. 
Instead, the interaction between sets of particles with a not so large difference between their velocities might be crucial for the generation of the coherent pulsar radio emission.
Another possibility would be that the essential interaction between the bunches takes place when the previously emitted bunch has already slowed down more than the successive one. 
This would imply that the difference between the minimum and maximum velocity is not as large anymore as it was right after their release. However, no known studies about a possible slow-down of the bunches are available so far.\\

Apart from the numerical predictions of the growth rates, all our analytical calculations are accompanied by PIC simulations that cover the same parameter space. The position of the dispersion branches is almost identical for all conducted simulations and the growth rate profiles have maxima for similar values of the wavenumber $k$.
The fits carried out in case of the simulation growth rates feature  a  high  precision  for  low  density  ratios  and  high temperatures, and they are reasonably accurate throughout the entire  beam  velocity  range. 
The agreement between simulation and analytical solutions corresponds to a co-validation of both approaches.
Yet,  they  deviate  with  decreasing temperature and increasing beam particle density.

The differences between PIC simulations, which solve the Vlasov-Maxwell
system of equations, and linear theory calculations based on a linearisation of the same system
were interpreted by \citet{Skoutnev2019} as being due to the 
enhanced PIC noise that might make the beams hotter than the assumed values, in this way weakening the linear theory predictions for the growth rates.
This effect could be especially significant for cold temperatures since the ratio between 
the enhanced (due to the noise) thermal fluctuations and beam temperature is large.
Yet, this effect does not equally apply to the strongly magnetised
plasma described by one-dimensional velocity distributions in our case.
Instead, the instability evolution, as it is quantified by the growth rates,
is not significantly influenced by the temperature uncertainty
as the growth rates are almost independent of the temperature
for the cold plasma limit, $\rho \rightarrow \infty$.
In addition, the growth rates are actually larger for simulations than the linear theory predictions.
This conclusion was also confirmed by two test simulations,
in which we increased either the number of particles per cell by a factor of eight
or the domain size by factor of ten. The resulting quantities fit within the corresponding uncertainty intervals as estimated in Table~\ref{tab:simulations}.

A different explanation of the observed deviations might apply to our case: As the differences between linear theory and simulations are found to be at large beam densities or lower temperatures, they might be caused by the different nature of the instability itself and 
the approximations that are considered in linear theory.
Indeed, streaming instabilities for large beam densities or lower temperatures
are usually within the regime known as reactive instabilities, 
characterised by a large wavenumber range and large growth rates \citep{Melrose1986}.
By contrast, high temperatures and lower beam densities favour the so-called kinetic 
or resonant
instabilities that feature a narrower unstable bandwidth (around the resonant phase speeds) and smaller growth rates.
We note that both are just different limits of the same instability.
Specifically, this explanation would be consistent with the larger growth rates and fractional bandwidths found in the simulations in Figures~\ref{fig:constrho1}-\ref{fig:rhobackground}. 
Assuming that our simulation results are not significantly influenced by numerical effects,
and in this way closer to reality than linear theory, we can estimate the origin of the
deviations found in the linear theory values to be higher-order effects not taken into account, in particular 
near the end of the linear stage of the instability evolution.
Since reactive instabilities for large beam densities or lower temperatures
have higher growth rates and more wave modes are excited (large bandwidth) than in the opposite regime, non-linear effects could modify the linear theory 
predictions due to possible large-amplitude waves partially captured 
during the linear fitting calculation in simulations.
In other words, there could be a quicker breakdown of the linear theory predictions in
the reactive instability limit than in the kinetic or resonant instability limit.
However, further investiagtions building on this are required in order to find out the exact cause 
of the disagreement between linear theory and simulations in this parameter regime.

\section{Conclusions}
\label{sec:conclusions}
We investigated one of the promising pulsar radio emission mechanisms that is based on the release of plasma bunches by computing the growth rates 
as a solution of the relativistic kinetic dispersion relation and by fully kinetic PIC simulations.
To this end, we carried out a parametric study, analysing the influence of the density ratio between beam and background, the beam and background temperatures, and the beam velocity. 
While some studies \citep{Melrose2017c,Rafat2019b} show that the growth rates from such an interaction are not high enough to consider this instability as a viable radio emission mechanism, we prove that for a specific parameter range, sufficiently large growth rates can be achieved.
For the instability to be efficient in the sense of being responsible for the pulsar radio emission generation, we found that the density ratio should be in the range $r_\mathrm{n} \geq 10^{-3}$, the inverse temperatures $\rho_0,\rho_1 > 1$, and the beam Lorentz factor $13 < \gamma_\mathrm{b} < 300$.

Moreover, we found a 
considerably good agreement between the analytical calculations (solutions of the linear dispersion relation) and PIC simulations. 
This was illustrated by a comparison between the growth rate profile as a function of the wave number.
In both cases, the growth rate maxima are very similar and have the same location in the $k$-space. We showed that the growth rates approach the expected growth rate values from the cold reactive beam instability in both cases. There are, however, small differences,
mostly in the cold plasma limit: The simulations show slightly larger growth rates than those of the linear 
theory predictions. This could be attributed to the quicker breakdown of linear theory in this parameter regime.

We also studied the transition between the regimes of a weak and strong beam
by investigating the dependence on its density ratio with respect to the background population, and the growth rate saturation as a function of the background and beam temperature.

Based on our findings, the two-stream instability should be further investigated for our selected parameters as a possible cause for pulsar radio emission. 
In particular, a comparison with observations is conceivable. A possible comparable quantity might consist in radio emission fluxes that can be obtained from the given calculated quantities.

\begin{acknowledgements}
We gratefully acknowledge the developers of the ACRONYM code
(\textit{Verein zur F\"orderung kinetischer Plasmasimulationen e.V.}),
and the financial support by the German Science Foundation (DFG) via the
projects MU-4255/1-1 and BU-777-17-1 as well as by 
the Czech Science Foundation (GACR) via the project 20-09922J.
This work was supported by The Ministry of Education, Youth and Sports from the Large Infrastructures for Research, Experimental Development and Innovations project ``e-Infrastructure CZ – LM2018140''.
Part of the simulations were carried out on the
HPC-Cluster of the Institute for Mathematics at the TU Berlin.
Computational resources were also supplied by the project "e-Infrastruktura CZ" (e-INFRA LM2018140) provided within the program Projects of Large Research, Development and Innovations Infrastructures.
The authors gratefully acknowledge the Gauss Centre for Supercomputing e.V. (\url{www.gauss-centre.eu}) for funding this project by providing computing time on the GCS Supercomputer SuperMUC-NG at Leibniz Supercomputing Centre (\url{www.lrz.de}), through the project pr27ta.
We thank the referee for his/her valuable comments to improve this paper.\\

The data sets used to reproduce the plots of this paper can be found at Zenodo: \url{https://zenodo.org/record/4049883#.X65Xy-1CdPY}
\end{acknowledgements}

%
%


\begin{thebibliography}{62}
\expandafter\ifx\csname natexlab\endcsname\relax\def\natexlab#1{#1}\fi

\bibitem[{Arendt \& Eilek(2002)}]{Arendt2002}
Arendt, P.~N. \& Eilek, J.~A. 2002, Astrophys. J., 581, 451

\bibitem[{Arons(1981)}]{Arons1981}
Arons, J. 1981, Astrophys. J., 248, 1099

\bibitem[{Asseo \& Melikidze(1998)}]{Asseo1998}
Asseo, E. \& Melikidze, G.~I. 1998, Mon. Not. R. Astron. Soc., 301, 59

\bibitem[{Baumjohann \& Treumann(1997)}]{Baum1997}
Baumjohann, W. \& Treumann, R. 1997, Basic Space Plasma Physics (Imperial
  College Press)

\bibitem[{Beskin(2018)}]{Beskin2018}
Beskin, V.~S. 2018, Uspekhi Fiz. Nauk, 188, 377

\bibitem[{Beskin {et~al.}(1993)Beskin, Gurevich, \& Istomin}]{Beskin1993}
Beskin, V.~S., Gurevich, S.~V., \& Istomin, Y.~N. 1993, {Physics of the pulsar
  magnetosphere} (Cambridge University Press)

\bibitem[{Blaskiewicz {et~al.}(1991)Blaskiewicz, Cordes, \&
  Wasserman}]{Blaskiewicz1991}
Blaskiewicz, M., Cordes, J.~M., \& Wasserman, I. 1991, The Astrophysical
  Journal, 370, 643

\bibitem[{Bret {et~al.}(2008)Bret, Gremillet, Benisti, \& Lefebvre}]{Bret2008}
Bret, A., Gremillet, L., Benisti, D., \& Lefebvre, E. 2008, Physical Review
  Letters, 100, 205008

\bibitem[{Bret {et~al.}(2010)Bret, Gremillet, \& Dieckmann}]{Bret2010}
Bret, A., Gremillet, L., \& Dieckmann, M.~E. 2010, Physics of Plasmas, 17,
  120501

\bibitem[{Buschauer \& Benford(1977)}]{Buschauer1977}
Buschauer, R. \& Benford, G. 1977, Mon. Not. R. Astron. Soc., 179, 99

\bibitem[{Cheng \& Ruderman(1977{\natexlab{a}})}]{Cheng1977c}
Cheng, A.~F. \& Ruderman, M.~A. 1977{\natexlab{a}}, Astrophys. J., 212, 800

\bibitem[{Cheng \& Ruderman(1977{\natexlab{b}})}]{Cheng1977b}
Cheng, A.~F. \& Ruderman, M.~A. 1977{\natexlab{b}}, Astrophys. J., 214, 598

\bibitem[{{Cole}(1997)}]{Cole1997}
{Cole}, J.~B. 1997, IEEE Transactions on Microwave Theory and Techniques, 45,
  991

\bibitem[{Cottrill {et~al.}(2008)Cottrill, Langdon, Lasinski, Lund, Molvig,
  Tabak, Town, \& Williams}]{Cottrill2008}
Cottrill, L.~A., Langdon, A.~B., Lasinski, B.~F., {et~al.} 2008, Physics of
  Plasmas, 15, 082108

\bibitem[{Diver \& Laing(2015)}]{Diver2015}
Diver, D.~A. \& Laing, E.~W. 2015, Physica Scripta, 90, 025602

\bibitem[{D’Angelo {et~al.}(2015)D’Angelo, Fedeli, Sgattoni, Pegoraro, \&
  Macchi}]{DAngelo2015}
D’Angelo, M., Fedeli, L., Sgattoni, A., Pegoraro, F., \& Macchi, A. 2015,
  Monthly Notices of the Royal Astronomical Society, 451, 3460–3467

\bibitem[{Egorenkov {et~al.}(1984)Egorenkov, Lominadze, \&
  Mamradze}]{Egorenkov1984}
Egorenkov, V.~D., Lominadze, D.~G., \& Mamradze, P.~G. 1984, Astrophysics
  (Engl. Transl.), 19, 426

\bibitem[{Eilek \& Hankins(2016)}]{Eilek2016}
Eilek, J. \& Hankins, T. 2016, J. Plasma Phys., 82, 635820302

\bibitem[{Esirkepov(2001)}]{Esikperov2001}
Esirkepov, T. 2001, Computer Physics Communications, 135, 144

\bibitem[{Gedalin {et~al.}(1998)Gedalin, Melrose, \& Gruman}]{Gedalin1998}
Gedalin, M., Melrose, D.~B., \& Gruman, E. 1998, Phys. Rev. E, 57, 3399

\bibitem[{Godfrey {et~al.}(1975)Godfrey, Newberger, \& Taggart}]{Godfrey1975a}
Godfrey, B.~B., Newberger, B.~S., \& Taggart, K.~A. 1975, IEEE Trans. Plasma
  Sci., 3, 60

\bibitem[{{Goldreich} \& {Julian}(1969)}]{Goldreich1969}
{Goldreich}, P. \& {Julian}, W.~H. 1969, Astrophysical Journal, 157, 869

\bibitem[{Hankins \& Eilek(2007)}]{Hankins2007}
Hankins, T.~H. \& Eilek, J.~A. 2007, Astrophys. J., 670, 693

\bibitem[{{Hewish} {et~al.}(1968){Hewish}, {Bell}, {Pilkington}, {Scott}, \&
  {Collins}}]{Hewish1968}
{Hewish}, A., {Bell}, S.~J., {Pilkington}, J.~D.~H., {Scott}, P.~F., \&
  {Collins}, R.~A. 1968, \nat, 217, 709

\bibitem[{Hinata(1976)}]{Hinata1976}
Hinata, S. 1976, 44, 389

\bibitem[{J{\"{u}}ttner(1911)}]{Juttner1911}
J{\"{u}}ttner, F. 1911, Ann. Phys., 339, 856

\bibitem[{Kilian {et~al.}(2012)Kilian, Burkart, \& Spanier}]{Kilian2012}
Kilian, P., Burkart, T., \& Spanier, F. 2012, in High Perform. Comput. Sci.
  Eng. '11, ed. W.~E. Nagel, D.~B. Kr{\"{o}}ner, \& M.~M. Resch (Berlin,
  Heidelberg: Springer Berlin Heidelberg), 5--13

\bibitem[{{Kramer} {et~al.}(2002){Kramer}, {Johnston}, \& {van
  Straten}}]{Kramer2002}
{Kramer}, M., {Johnston}, S., \& {van Straten}, W. 2002, Monthly Notices of the
  Royal Astronomical Society, 334, 523

\bibitem[{Kärkkäinen {et~al.}(2006)Kärkkäinen, Gjonaj, Lau, \&
  Weiland}]{Karkkainen2006}
Kärkkäinen, M., Gjonaj, E., Lau, T., \& Weiland, T. 2006, Proc. International
  Computational Accelerator Physics Conference

\bibitem[{Lesch {et~al.}(1998)Lesch, Jessner, Kramer, \& Kunzl}]{Lesch98}
Lesch, H., Jessner, A., Kramer, M., \& Kunzl, T. 1998, Astronomy and
  Astrophysics, 332

\bibitem[{Levenberg(1944)}]{Levenberg1944}
Levenberg, K. 1944, 11, 164

\bibitem[{{Liu} {et~al.}(2019){Liu}, {Young}, {Wharton}, {Blackburn},
  {Cappallo}, {Chatterjee}, {Cordes}, {Crew}, {Desvignes}, {Doeleman},
  {Eatough}, {Falcke}, {Goddi}, {Johnson}, {Johnston}, {Karuppusamy}, {Kramer},
  {Matthews}, {Ransom}, {Rezzolla}, {Rottmann}, {Tilanus}, \&
  {Torne}}]{Liu2019}
{Liu}, K., {Young}, A., {Wharton}, R., {et~al.} 2019, The Astrophysical Journal
  Letters, 885, L10

\bibitem[{{Lu} {et~al.}(2020){Lu}, {Kilian}, {Guo}, {Li}, \& {Liang}}]{Lu2020}
{Lu}, Y., {Kilian}, P., {Guo}, F., {Li}, H., \& {Liang}, E. 2020, Journal of
  Computational Physics, 413, 109388

\bibitem[{Luo \& Melrose(2001)}]{Luo2001}
Luo, Q. \& Melrose, D.~B. 2001, Mon. Not. R. Astron. Soc., 325, 187

\bibitem[{López {et~al.}(2014)López, Muñoz, Viñas, \& Valdivia}]{Lopez2014}
López, R.~A., Muñoz, V., Viñas, A.~F., \& Valdivia, J.~A. 2014, Physics of
  Plasmas, 21, 032102

\bibitem[{López {et~al.}(2015)López, Navarro, Moya, F.~Viñas, Araneda,
  Muñoz, \& Valdivia}]{Lopez2015}
López, R.~A., Navarro, R.~E., Moya, P.~S., {et~al.} 2015, The Astrophysical
  Journal, 810, 103

\bibitem[{Marquardt(1963)}]{Marquardt1963}
Marquardt, D.~W. 1963, Journal of the Society for Industrial and Applied
  Mathematics, 11, 431–441

\bibitem[{{Melrose}(1978)}]{Melrose1978}
{Melrose}, D.~B. 1978, The Astrophysical Journal, 225, 557

\bibitem[{Melrose(1986)}]{Melrose1986}
Melrose, D.~B. 1986, {Instabilities in Space and Laboratory Plasmas}
  (Cambridge: Cambridge University Press)

\bibitem[{Melrose(2017)}]{Melrose2017}
Melrose, D.~B. 2017, Reviews of Modern Plasma Physics, 1, 5

\bibitem[{Melrose \& Gedalin(1999)}]{Melrose1999}
Melrose, D.~B. \& Gedalin, M.~E. 1999, Astrophys. J., 521, 351

\bibitem[{Melrose \& Rafat(2017)}]{Melrose2017c}
Melrose, D.~B. \& Rafat, M.~Z. 2017, in IOP Conf. Ser. 932, 012011

\bibitem[{{Petrova}(2009)}]{Petrova2009}
{Petrova}, S.~A. 2009, Monthly Notices of the Royal Astronomical Society, 395,
  1723

\bibitem[{Philippov {et~al.}(2020)Philippov, Timokhin, \&
  Spitkovsky}]{Philippov2020}
Philippov, A., Timokhin, A., \& Spitkovsky, A. 2020, Phys. Rev. Lett., 124,
  245101

\bibitem[{Powell(1964)}]{Powell1964}
Powell, M. J.~D. 1964, The Computer Journal, 7, 155

\bibitem[{Rafat {et~al.}(2019{\natexlab{a}})Rafat, Melrose, \&
  Mastrano}]{Rafat2019}
Rafat, M.~Z., Melrose, D.~B., \& Mastrano, A. 2019{\natexlab{a}}, J. Plasma
  Phys., 85, 905850305

\bibitem[{Rafat {et~al.}(2019{\natexlab{b}})Rafat, Melrose, \&
  Mastrano}]{Rafat2019b}
Rafat, M.~Z., Melrose, D.~B., \& Mastrano, A. 2019{\natexlab{b}}, J. Plasma
  Phys., 85, 905850603

\bibitem[{Rahaman {et~al.}(2020)Rahaman, Mitra, \& Melikidze}]{Rahaman2020}
Rahaman, S.~M., Mitra, D., \& Melikidze, G.~I. 2020, Monthly Notices of the
  Royal Astronomical Society, 497, 3953

\bibitem[{Ruderman \& Sutherland(1975)}]{Ruderman1975}
Ruderman, M.~A. \& Sutherland, P.~G. 1975, Astrophys. J., 196, 51

\bibitem[{Shalaby {et~al.}(2017)Shalaby, Broderick, Chang, Pfrommer, Lamberts,
  \& Puchwein}]{Shalaby2017}
Shalaby, M., Broderick, A.~E., Chang, P., {et~al.} 2017, The Astrophysical
  Journal, 841, 52

\bibitem[{Shalaby {et~al.}(2018)Shalaby, Broderick, Chang, Pfrommer, Lamberts,
  \& Puchwein}]{Shalaby2018}
Shalaby, M., Broderick, A.~E., Chang, P., {et~al.} 2018, The Astrophysical
  Journal, 859, 45

\bibitem[{Shukla {et~al.}(2015)Shukla, Das, \& Patel}]{Shukla2015}
Shukla, C., Das, A., \& Patel, K. 2015, Physics of Plasmas, 22, 112118

\bibitem[{Silva {et~al.}(2003)Silva, Fonseca, Tonge, Dawson, Mori, \&
  Medvedev}]{Silva2003}
Silva, L.~O., Fonseca, R.~A., Tonge, J.~W., {et~al.} 2003, The Astrophysical
  Journal, 596, L121–L124

\bibitem[{Skoutnev {et~al.}(2019)Skoutnev, Hakim, Juno, \&
  TenBarge}]{Skoutnev2019}
Skoutnev, V., Hakim, A., Juno, J., \& TenBarge, J.~M. 2019, The Astrophysical
  Journal, 872, L28

\bibitem[{Stenson {et~al.}(2017)Stenson, Horn-Stanja, Stoneking, \&
  Pedersen}]{Stenson2017}
Stenson, E.~V., Horn-Stanja, J., Stoneking, M.~R., \& Pedersen, T.~S. 2017, J.
  Plasma Phys., 83, 595830106

\bibitem[{Sturrock(1958)}]{Sturrock1958}
Sturrock, P.~A. 1958, Phys. Rev., 112, 1488

\bibitem[{Sturrock(1971)}]{Sturrock1971}
Sturrock, P.~A. 1971, Astrophys. J., 164, 529

\bibitem[{Tautz {et~al.}(2007)Tautz, Sakai, \& Lerche}]{Tautz2007}
Tautz, R.~C., Sakai, J.-I., \& Lerche, I. 2007, Astrophysics and Space Science,
  310, 159–167

\bibitem[{Ursov \& Usov(1988)}]{Ursov1988}
Ursov, V. \& Usov, V. 1988, Astrophysics and Space Science, 140, 325

\bibitem[{Usov(1987)}]{Usov1987}
Usov, V.~V. 1987, Astrophys. J., 320, 333

\bibitem[{{Usov}(2002)}]{Usov2002}
{Usov}, V.~V. 2002, in Neutron Stars, Pulsars, and Supernova Remnants, ed.
  W.~{Becker}, H.~{Lesch}, \& J.~{Tr{\"u}mper}, 240

\bibitem[{Weatherall(1994)}]{Weatherall1994}
Weatherall, J.~C. 1994, Astrophys. J., 428, 261

\end{thebibliography}

\end{document}